\documentstyle[prb,twocolumn,aps,eqsecnum]{revtex}
\begin{document}

\twocolumn[\hsize\textwidth\columnwidth\hsize\csname
@twocolumnfalse\endcsname
\title{ Liquid $\bf ^4$He near the superfluid transition in the presence of a 
heat current and gravity }

\author{ Rudolf Haussmann }

\address{ Sektion Physik, Universit\"at M\"unchen, Theresienstrasse 37, 
D-80333 M\"unchen, Germany }

\date{ submitted to Physical Review B, November 4, 1998, accepted }

\maketitle

\begin{abstract}
The effects of a heat current and gravity in liquid $^4$He near the superfluid 
transition are investigated for temperatures above and below $T_\lambda$. We 
present a renormalization-group calculation based on model {\it F\,} for the 
Green's function in a self-consistent approximation which in quantum 
many-particle theory is known as the Hartree approximation. The approach can 
handle the average order parameter $\langle\psi\rangle=0$ above and below 
$T_\lambda$ and includes effects of vortices. We calculate the thermal 
conductivity $\lambda_{\rm T}(\Delta T,Q)$ and the specific heat 
$C(\Delta T,Q)$ for all temperature differences $\Delta T=T-T_\lambda$ and 
heat currents $Q$ in the critical regime. Furthermore, we calculate the 
temperature profile $T(z)$. Below $T_\lambda$ we find a second correlation 
length $\xi_1 \sim Q^{-1}(T_\lambda-T)^{+\nu}$ which describes the dephasing 
of the order-parameter field due to vortices. We find dissipation and mutual 
friction of the superfluid-normal fluid counterflow and calculate the 
Gorter-Mellink coefficient $A$. We compare our theoretical results with 
recent experiments.
\end{abstract}

\vskip2pc]

\section{Introduction} \label{S01}
Gravity and a heat current $Q$ are two sources which influence the 
superfluid transition of liquid $^4$He at $T_\lambda \approx 2\,{\rm K}$ and 
cause inhomogeneities in the system. On earth gravity implies a pressure
variation $P=P(z)$ where $z$ is the altitude coordinate. Since the superfluid
transition temperature $T_\lambda=T_\lambda(P)$ is pressure dependent, 
$T_\lambda(z)=T_\lambda(P(z))$ depends on the altitude coordinate $z$ with
the gradient \cite{A0} $\partial T_\lambda/\partial z= +1.273\ \mu{\rm K/cm}$.
On the other hand a nonzero heat current $Q$ drives the system away from
equilibrium. A temperature gradient $\bbox{\nabla}T$ is created which implies
that the temperature $T$ is space dependent. We assume that the heat current
$Q$ is homogeneous and flows vertically (parallel to the $z$ axis) so that 
the temperature $T(z)$ depends on the $z$ coordinate only. 

The local properties of the system are determined by three parameters, 
the local temperature difference $\Delta T(z)= T(z)-T_\lambda(z)$, the heat 
current $Q$, which is related to $\bbox{\nabla}T$, and gravity $g$, which is 
related to $\bbox{\nabla}T_\lambda$. The point $(\Delta T,Q,g)=(0,0,0)$ is a 
critical point related to the superfluid transition. This means that in 
thermal equilibrium ($Q=0$) and in microgravity ($g=0$) the system shows a 
second-order phase transition at $T=T_\lambda$ from the normal-fluid to
the superfluid state. Usually, gravity is negligible except for very small
heat currents $Q$ and for very small $\Delta T$, i.e.\ very close to 
$T_\lambda$. Since on earth the gravity acceleration $g=9.81\ \mbox{m/s}^2$ 
is a fixed quantity, in most cases the $\Delta T$-$Q$ plane is considered as 
the phase diagram.

Liquid $^4$He close to $T_\lambda$ in the presence of a heat current $Q$ has
been investigated theoretically \cite{O1,HD1,HD2} and experimentally
\cite{DAS,LA1,MD,MM}. In the $\Delta T$-$Q$ plane a line of critical
temperatures is found which separates superfluid from normal-fluid helium. 
A nonzero heat current $Q$ implies that the superfluid transition temperature
$T_\lambda$ is shifted to lower temperatures by $\Delta T_\lambda(Q)$. For
small heat currents $Q$ in the critical regime the theory \cite{O1,HD2}
predicts the shift $\Delta T_\lambda(Q) \sim -Q^x$ with the exponent 
$x=1/2\nu = 0.745$. In the experiments \cite{DAS,EP,LP,BL} a depression 
$\Delta T_{\rm c}(Q)$ of the superfluid transition has been observed which
agrees qualitatively with the theory, but not quantitatively. For small $Q$
in the critical regime the shift $\Delta T_{\rm c}(Q)\sim -Q^{0.81}$ has been
found \cite{DAS}. While the exponents do not agree, the experimentally 
observed shift $\Delta T_{\rm c}(Q)$ is larger than the theoretically 
calculated $\Delta T_\lambda(Q)$. 

While for $\Delta T\gtrsim \Delta T_\lambda(Q)$ the helium is normal fluid
and for $\Delta T \leq \Delta T_{\rm c}(Q)$ it is superfluid, in a recent
experiment Liu and Ahlers \cite{LA1} found a new dissipative region for
temperatures $\Delta T$ in the interval $\Delta T_{\rm c}(Q) < \Delta T <
\Delta T_\lambda(Q)$. This observation indicates that at a finite heat 
current $Q$ the transition from normal fluid to superfluid helium may 
possibly happen in two steps with two transition temperatures $\Delta 
T_\lambda(Q)$ and $\Delta T_{\rm c}(Q)$ (relative to the equilibrium 
transition temperature $T_\lambda$). While the upper $\Delta T_\lambda(Q)$ 
may be identified by the theoretical prediction, the lower $\Delta T_{\rm c}
(Q)$ agrees with the shift of Ref.\ \onlinecite{DAS}. In a similar experiment 
performed by Murphy and Meyer \cite{MM} also two transition temperatures were 
found. While the values of $\Delta T_\lambda(Q)$ and $\Delta T_{\rm c}(Q)$ 
contain errors, the difference $\Delta T_\lambda(Q) - \Delta T_{\rm c}(Q)$
is quite well reproduced by the latter experiment. 

Heat-transport phenomena in liquid $^4$He close to $T_\lambda$ are described
by model {\it F\,} of Halperin, Hohenberg, and Siggia \cite{HH} which is a 
model for the critical and hydrodynamic slow variables including fluctuations. 
Most theoretical investigations are based on this model. In the normal-fluid 
region for temperatures $T$ above $T_\lambda$ the heat is transported 
diffusively driven by the temperature gradient $\bbox{\nabla}T$. In linear 
response the heat current is ${\bf Q}=-\lambda_{\rm T} \bbox{\nabla}T$ where 
$\lambda_{\rm T}$ is the thermal conductivity. For infinitesimal $Q$ and zero 
gravity the thermal conductivity $\lambda_{\rm T}$ has been calculated within 
model {\it F\,} in two-loop order \cite{D1}. Critical fluctuations, which are 
taken into account by the renormalization-group (RG) theory, imply a strong 
enhancement of $\lambda_{\rm T}$ close to $T_\lambda$. For infinitesimal $Q$ 
and zero gravity the thermal conductivity $\lambda_{\rm T}$ diverges in the 
limit $T\rightarrow T_\lambda$, or more precisely in the limit $(\Delta T,Q,g)
\rightarrow (0,0,0)$. The RG theory has been extended \cite{HD1} to calculate 
$\lambda_{\rm T}$ for nonzero heat currents $Q$ but without gravity. It turns 
out that for finite $Q$ close to $T_\lambda$ the heat transport becomes 
nonlinear which means that $\lambda_{\rm T}$ becomes $Q$-dependent. It has 
been shown \cite{HD1} that $\lambda_{\rm T}$ remains finite in this case even 
for $T=T_\lambda$.

On the other hand, in the superfluid region where $\Delta T(z)< \Delta 
T_{\rm c}(Q)$ the heat is transported convectively nearly without friction 
according to the two-fluid model \cite{L1} by the superfluid-normal fluid 
counterflow. In this case the temperature gradient $\bbox{\nabla}T$ is nearly 
zero indicating a nearly infinite thermal conductivity $\lambda_{\rm T}$. 
Mutual friction between the superfluid and the normal-fluid component and
dissipation of the heat current occur only by creation of vortices, which 
however is a small effect. Nevertheless, mutual friction in the superfluid
state has been measured as early as 1949 by Gorter and Mellink \cite{GM}. 
For the mutual friction force the ansatz $f=A\rho_{\rm n}\rho_{\rm s}
(v_{\rm s}-v_{\rm n})^3$ was made \cite{GM} with a temperature dependent
coefficient $A$, the so called Gorter-Mellink coefficient. This ansatz is
related to a turbulent superfluid flow \cite{V1} and implies a temperature 
gradient $\bbox{\nabla}T\sim -Q^3$, so that the thermal conductivity is 
$\lambda_{\rm T}\sim Q^{-2}$. More recently, the temperature gradient 
$\bbox{\nabla}T$ due to mutual friction in superfluid $^4$He was measured
directly by Baddar et al.\ \cite{BA} in the critical regime close to 
$T_\lambda$. This experiment confirms the ansatz by Gorter and Mellink 
qualitatively but with a slightly different exponent in the $Q$ dependence 
or $v_{\rm s}-v_{\rm n}$ dependence. 

In the intermediate region a crossover between the two heat-transport 
mechanisms happens. Since the system is spatially inhomogeneous this 
crossover happens also spatially and implies an interface between superfluid
and normal-fluid helium located at a certain $z_0$. Onuki \cite{O1} 
investigated this interface by solving the model-{\it F\,} equations in 
mean-field approximation, where critical fluctuations are taken into account 
by scaling theory. He calculated the temperature profile $T(z)$ and the 
order-parameter profile $\psi(z)$. While in the normal-fluid region the
temperature profile has a finite gradient related to a finite thermal
conductivity $\lambda_{\rm T}$, in the superfluid region the temperature
profile is absolutely flat and the gradient is zero, so that mutual friction
is not included and $\lambda_{\rm T}$ is infinite.

The previous RG theories \cite{HD1,HD2} were constructed as perturbation 
theories starting with mean-field solutions of the model {\it F\,} equations.
While in the normal-fluid region it is \cite{HD1} $\langle\psi\rangle=0$, in 
the superfluid region a plane wave order parameter $\langle\psi\rangle=\eta\,
e^{ikz}$ was assumed \cite{HD2}. Consequently, the temperature profile was
found to be flat in the superfluid region so that mutual friction and 
dissipation by vortex creation are not included in the previous RG theories. 
In this paper we extend the RG theory in two respects. First, we include 
gravity. Secondly, we include mutual friction and dissipation in the
superfluid region.

Strong fluctuations of the phase of the order-parameter field $\psi$ can
imply that the average order parameter is $\langle\psi\rangle=0$ even below 
$T_\lambda$. This fact is well known for systems of finite size. Here for 
nonzero $Q$ the vortices cause sufficiently strong phase fluctuations so 
that $\langle\psi\rangle=0$ not only above but also below $T_\lambda$. 
For this reason, in the nonequilibrium state with a nonzero heat current $Q$ 
we may not start a perturbation theory with a nonzero mean-field order 
parameter which implies a nonzero $\langle\psi\rangle$. Rather we need a new 
approach which can handle $\langle\psi\rangle=0$ above and below $T_\lambda$. 
In this paper we present a self-consistent approximation for the Green's 
function which can do this and which in quantum many-particle theory is known 
as the Hartree approximation \cite{FW}. 

The paper is organized as follows. In Sec.\ \ref{S02} we describe the model
and the necessary field theoretic tools, i.e.\ the Feynman rules. In 
Sec.\ \ref{S03} the idea of our approach and the calculations of the Green's 
function and the effective parameters are presented. The 
renormalization-group theory is applied in Sec.\ \ref{S04} to include the 
critical fluctuations. The thermal conductivity and the temperature profile
are calculated in Secs.\ \ref{S05} and \ref{S06}, while in Secs.\ \ref{S07}
and \ref{S08} the correlation lengths, the entropy, and the specific heat are 
considered. We compare our results with experiments and investigate the 
influence of gravity. In Sec.\ \ref{S09} we discuss dissipation and mutual 
friction for superfluid helium below $T_\lambda$. We show that our approach 
reproduces the ansatz of Gorter and Mellink \cite{GM} for the mutual friction 
force and calculate the Gorter-Mellink coefficient $A$. The idea of the 
approach and part of the results have been published already in a rapid
communication \cite{H1}.

\section{The model and Feynman rules} \label{S02}
Dynamic critical and heat transport phenomena in liquid $^4$He close to
$T_\lambda$ are well described by model {\it F\,} which is given \cite{HH} by 
the Langevin equations for the order parameter $\psi({\bf r},t)$ and the 
entropy variable $m({\bf r},t)$:
\begin{eqnarray}
{\partial \psi\over \partial t} &=& -2\Gamma_0 {\delta H\over \delta \psi^*}
+ ig_0 \psi {\delta H\over \delta m} + \theta_\psi \ ,
\label{A01} \\
{\partial m\over \partial t} &=& \lambda_0 \bbox{\nabla}^2 {\delta H\over
\delta m} - 2g_0 {\rm Im} \Bigl( \psi^* {\delta H\over \delta \psi^*} 
\Bigr) + \theta_m \ , \label{A02}
\end{eqnarray}
where
\begin{eqnarray}
H &=& \int d^d r \bigl[ {\textstyle\frac{1}{2}} \tau_0(z) \vert\psi\vert^2 
+ {\textstyle\frac{1}{2}} \vert\bbox{\nabla}\psi\vert^2 + \tilde u_0 
\vert\psi\vert^4 \nonumber\\
&&\hskip1.1cm + {\textstyle\frac{1}{2}} \chi_0^{-1} m^2 + \gamma_0 m 
\vert\psi\vert^2 - h_0 m \bigr]  \label{A03} 
\end{eqnarray}
is the free energy functional and $\theta_\psi$ and $\theta_m$ are Gaussian
stochastic forces which incorporate the fluctuations. The heat current $Q$ is
imposed by boundary conditions. The gravity is included via the temperature
parameter $\tau_0(z)$ in (\ref{A03}) which is related to $T_\lambda(z)$ and 
depends linearly on the altitude $z$. Usually, the model is treated by 
field-theoretic means. The perturbation theory in terms of Feynman diagrams
is generated by the Janssen-De Dominicis functional integral \cite{JD}
\begin{equation}
Z = \int D\psi D\tilde\psi \, Dm D\tilde m \ \exp\{J\}
\label{A04}
\end{equation}
with the functional
\begin{eqnarray}
J &=& \int d^d r \int dt \, \Bigl\{ \lambda_0 (\bbox{\nabla}\tilde m)^2
+ \Gamma_0^\prime \vert\tilde\psi\vert^2  \nonumber\\
&& - {1\over 2} \tilde\psi^* \Bigl[ {\partial\psi\over\partial t} +
2\Gamma_0 {\delta H\over\delta \psi^*} - i g_0 \psi {\delta H\over\delta m}
\Bigr]  \nonumber\\
&& - {1\over 2} \tilde\psi \Bigl[ {\partial\psi^*\over\partial t} +
2\Gamma_0^* {\delta H\over\delta \psi} + i g_0 \psi^* {\delta H\over\delta m}
\Bigr]  \nonumber\\
&& - \tilde m \Bigl[ {\partial m\over\partial t} - \lambda_0 \bbox{\nabla}^2
{\delta H\over\delta m} + 2 g_0 {\rm Im} \Bigl( \psi^* {\delta H\over\delta
\psi^*} \Bigr)  \Bigr]  \Bigr\} \ . 
\label{A05}
\end{eqnarray}
Here $\tilde\psi$ and $\tilde m$ are auxiliary fields \cite{MSR} which are
needed for a proper construction of the perturbation theory. While in the
previous theories \cite{HD1,HD2} only the free Green's functions were needed,
here we will evaluate the self energy in leading order. For this reason, we
briefly describe the Feynman rules by which the diagrams and the terms of
the perturbation series are constructed. We decompose the fields into a 
mean-field and a fluctuating contribution according to $\psi = \psi_{\rm mf}
+ \delta\psi$ and $m=m_{\rm mf} + \delta m$ where $\psi_{\rm mf}$ and 
$m_{\rm mf}$ are solutions of the model-{\it F\,} equations (\ref{A01}) and
(\ref{A02}) without the stochastic forces. Here we assume a zero mean-field
order parameter $\psi_{\rm mf}=0$ above and below $T_\lambda$, so that $\psi=
\delta\psi$. The boundary conditions, which imply the heat current, require
\begin{equation}
m_{\rm mf}(z) = - q (\chi_0/\lambda_0) (z-z_0)
\label{A06}
\end{equation}
where $q$ is the entropy current related to the heat current $Q$ in physical
units by $q = Q/k_{\rm B} T_\lambda$. We decompose the functional $J$ in 
powers of the fluctuating fields according to
\begin{equation}
J = J_2 + J_3 + J_4 \ . 
\label{A07}
\end{equation}
The quadratic term is given by
\begin{eqnarray}
J_2 &=& \int d^d r \int dt \, \bigl[ \lambda_0 (\bbox{\nabla}\tilde m)^2
+ \Gamma_0^\prime \vert\tilde\psi\vert^2  \nonumber\\
&& - \textstyle{1\over 2} \tilde\psi^* L_0 \psi 
- \textstyle{1\over 2} \psi^* L_0^+ \tilde\psi - \tilde m N_0 \delta m \bigr]
\label{A08}
\end{eqnarray}
with the differential operators 
\begin{eqnarray}
L_0 &=& \partial_t + \Gamma_0 [\tau_0(z) - \bbox{\nabla}^2 + 2 \gamma_0
m_{\rm mf}(z) ] \nonumber\\ 
&&\hspace{1cm} - i(g_0/\chi_0) [m_{\rm mf}(z) - \chi_0 h_0]
\label{A09}
\end{eqnarray}
and
\begin{equation}
N_0 = \partial_t - (\lambda_0/\chi_0) \bbox{\nabla}^2 \ .
\label{A10}
\end{equation}
The third and fourth order terms are given by
\begin{eqnarray}
J_3 &=& \int d^d r \int dt \, \bigl\{ \tilde m [ \lambda_0\gamma_0 
\bbox{\nabla}^2 (\psi^*\psi) + g_0 \bbox{\nabla} {\rm Im}(\psi^* \bbox{\nabla}
\psi) ]  \nonumber\\
&& \hspace{2cm} - b_3\, \delta m\, \tilde\psi^* \psi - b_3^*\, \delta m\, 
\psi^* \tilde\psi \bigr\} 
\label{A11}
\end{eqnarray}
and 
\begin{equation}
J_4 = \int d^d r \int dt \, \bigl\{ -b_4\, \tilde\psi^* \psi^* \psi\, \psi  
- b_4^*\, \psi^* \psi^* \psi\, \tilde \psi \bigr\} \ ,  
\label{A12}
\end{equation}
respectively, where 
\begin{eqnarray}
b_3 = \Gamma_0 \gamma_0 - i g_0 / 2\chi_0 \ ,
\hspace{0.5cm} b_4 = 2 \Gamma_0 \tilde u_0 - i g_0 \gamma_0 /2
\label{A13}
\end{eqnarray}
are complex parameters. While the free Green's functions are obtained from
the quadratic term $J_2$, the interaction vertices are obtained from $J_3$
and $J_4$. In order to obtain a compact notation of the Feynman rules we 
combine the fields into vectors
\begin{equation}
(\Psi_\alpha) = \pmatrix{ \tilde\psi \cr \psi \cr } \ , \hspace{0.5cm}
(M_\alpha) = \pmatrix{ \tilde m \cr \delta m \cr } 
\label{A14}
\end{equation}
where the index $\alpha = 1,2$ distinguishes between fields with and without
tilde. Then the several contributions of the functional $J$ can be written as
\begin{eqnarray}
J_2 &=& - \! \int \! d^d r \int \! dt \bigl[ \Psi_\alpha^* K_{0,\alpha\beta} 
\Psi_\beta + \textstyle{1\over 2} M_\alpha E_{0, \alpha\beta} M_\beta 
\bigr] ,  \label{A15}  \\
J_3 &=& - \int d^d r \int dt \, B_{3,\alpha\beta\gamma} M_\alpha \Psi_\beta^*
\Psi_\gamma  \ , \label{A16}  \\
J_4 &=& - {\textstyle{1\over 2}} \int d^d r \int dt \, B_{4,\alpha\beta,
\gamma\delta} (\Psi_\alpha^* \Psi_\beta) (\Psi_\gamma^* \Psi_\delta)  \ . 
\label{A17}
\end{eqnarray}
Here
\begin{eqnarray}
(K_{0,\alpha\beta}) &=& \pmatrix{ -\Gamma_0^\prime & {1\over 2} L_0 \cr 
{1\over 2} L_0^+ & 0 \cr} \ , \label{A18}  \\
(E_{0,\alpha\beta}) &=& \pmatrix{ 2\lambda_0 \bbox{\nabla}^2 & N_0 \cr 
N_0^+ & 0 \cr} \label{A19}
\end{eqnarray}
are $2\times 2$ matrices with differential operators as the elements. The 
free Green's functions are obtained by inverting these matrices according to
\begin{eqnarray}
G_{0,\alpha\beta}({\bf r},t;{\bf r}^\prime,t^\prime) &=& 
\langle \Psi_\alpha({\bf r},t) \Psi_\beta^*({\bf r}^\prime,t^\prime) \rangle_0
\nonumber\\
&=& K^{-1}_{0,\alpha\beta} \, \delta({\bf r} - {\bf r}^\prime) 
\, \delta(t-t^\prime) \ ,  \label{A20}  \\
D_{0,\alpha\beta}({\bf r},t;{\bf r}^\prime,t^\prime) &=& 
\langle M_\alpha({\bf r},t) M_\beta({\bf r}^\prime,t^\prime) \rangle_0
\nonumber\\
&=& E^{-1}_{0,\alpha\beta} \, \delta({\bf r} - {\bf r}^\prime) 
\, \delta(t-t^\prime) \ .  \label{A21}
\end{eqnarray}
The higher-rank tensors $B_{3,\alpha\beta\gamma}$ and $B_{4,\alpha\beta,
\gamma\delta}$, which describe the interactions between the fields, are
obtained by comparing (\ref{A16}) and (\ref{A17}) with (\ref{A11}) and
(\ref{A12}), respectively. While $B_{3,\alpha\beta\gamma}$ contains 
differential operators, $B_{4,\alpha\beta,\gamma\delta}$ is symmetrized with
respect to interchange of the index pairs $(\alpha,\beta)$ and $(\gamma,
\delta)$.

\begin{figure}
\vspace*{6.1cm}
\includegraphics{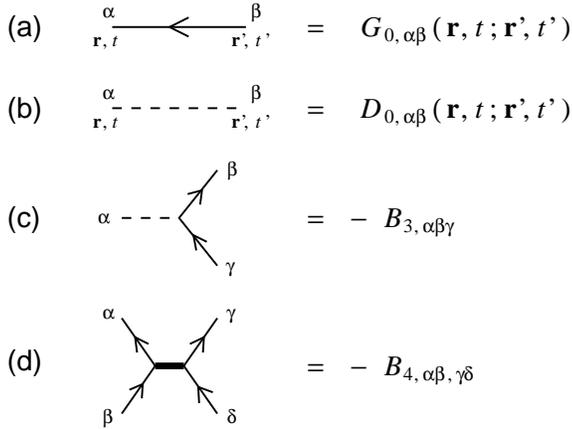}
\caption{The elements for constructing the Feynman diagrams: (a) free
$\psi$-field Green's function, (b) free $m$-field Green's function, (c)
three vertex, and (d) four vertex.}
\label{F01}
\end{figure}

Now, from (\ref{A15})-(\ref{A21}) the Feynman rules are obtained easily.
In Fig.\ \ref{F01} the elements for constructing the Feynman diagrams are
shown. The free $\psi$-field Green's function $G_{0,\alpha\beta}({\bf r},t;
{\bf r}^\prime,t^\prime)$ is identified by a directed solid line (Fig.\  
\ref{F01}a). The free $m$-field Green's function $D_{0,\alpha\beta}
({\bf r},t;{\bf r}^\prime,t^\prime)$ is identified by a dashed line 
(Fig.\ \ref{F01}b). The $m$-field interacts with the $\psi$ fields by the
three vertex shown in Fig.\ \ref{F01}c. Furthermore, the $\psi$ fields 
interact with each other by the four vertex shown in Fig.\ \ref{F01}d. 
For each three or four vertex an integration $\int d^d r \int dt$ must be 
performed. Further rules are applied as usual in field theory. The 
perturbation series is obtained as the sum of all possible Feynman diagrams 
which can be constructed from the elements shown in Fig.\ \ref{F01}.

\section{The unrenormalized Green's function in Hartree approximation} 
\label{S03}
To obtain physical quantities we must first calculate the $\psi$-field
Green's function which is defined by
\begin{equation}
G_{\alpha\beta}({\bf r},t;{\bf r}^\prime,t^\prime) = \langle 
\Psi_\alpha({\bf r},t) \Psi_\beta^*({\bf r}^\prime,t^\prime) \rangle \ .
\label{B01}
\end{equation}
This Green's function can be expressed via the Dyson equation 
\begin{equation}
G^{-1} = G_0^{-1} - \Sigma
\label{B02}
\end{equation}
in terms of the self energy $\Sigma_{\alpha\beta}({\bf r},t;{\bf r}^\prime,
t^\prime)$. The perturbation series of the self energy is given by the sum
of all irreducible Feynman diagrams with two amputated external solid lines 
which do not fall into pieces if any internal solid line is cut. A similar
Dyson equation exists also for the $m$-field Green's function $D_{\alpha\beta}
({\bf r},t;{\bf r}^\prime,t^\prime)$. However, for our calculations we do not
need the latter Green's function explicitly. 

In order to account for effects beyond the perturbation theory we must resum
the Feynman diagrams partially in an appropriate way. In the critical regime
close to a second-order phase transition, infrared singularities occur which
must be resummed by renormalization and application of the 
renormalization-group theory. For model {\it F\,} the RG theory was elaborated 
up to two-loop order by Dohm \cite{D1}. All the renormalized coupling 
parameters depending on a RG flow parameter were determined \cite{D1} by 
adjusting the superfluid density, the specific heat, and the thermal 
conductivity to the respective experimental data. Thus, model {\it F\,} can be
used for explicit calculations of physical quantities in the critical regime 
near $T_\lambda$ without any (further) adjustable parameters.

\begin{figure}
\vspace*{3.4cm}
\includegraphics{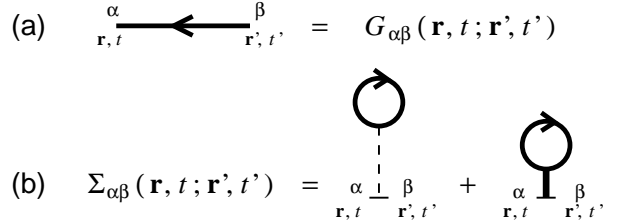}
\caption{(a) The exact $\psi$-field Green's function is identified by a thick
solid directed line. (b) The self energy in Hartree approximation.}
\label{F02}
\end{figure}

Here we first apply an additional resummation which is well known in quantum
many-particle physics \cite{LW,DM,AGD}. We resum with respect to all 
self-energy subdiagrams so that the perturbation series becomes self 
consistent with respect to the $\psi$-field Green's function $G$. This means
that now the solid lines are thick and identified by the exact Green's 
function $G$ as shown in Fig.\ \ref{F02}a. To avoid multiple counting of 
diagrams, only the irreducible diagrams are included in the perturbation 
series, which do not contain self-energy subdiagrams or equivalently which
do not fall into pieces if any two of the internal thick solid lines are cut.
The resummation was used first by Luttinger and Ward \cite{LW} and was
formulated in terms of a Legendre transformation by De Dominicis and Martin
\cite{DM}. By truncating the self-consistent perturbation series of the
self energy $\Sigma$ the conserving approximation of Baym and Kadanoff
\cite{BK,B1} is obtained. 

Here, we approximate the self energy $\Sigma$ by including only the tadpole 
diagrams as shown in Fig.\ \ref{F02}b. This approximation is equivalent to 
the Hartree approximation in quantum mechanics \cite{FW}. The Dyson equation 
(\ref{B02}) together with the self energy in Fig.\ \ref{F02}b are 
self-consistent equations which enable an explicit calculation of the 
unrenormalized Green's function $G$. However, since we consider liquid $^4$He 
in the critical regime near $T_\lambda$, a second resummation is necessary: 
the self-consistent perturbation series and hence the Hartree approximation 
must be modified by renormalization and application of the RG theory. 

We have several reasons to believe that the Hartree approximation combined 
with the RG theory is successful for model {\it F\,} above and below 
$T_\lambda$ where the order parameter is always $\langle\psi\rangle=0$. 
Moreover, we will show that the approximation includes vortices and mutual 
friction so that it may be a possible approach to describe the dissipation in 
the superfluid state observed in the experiments \cite{LA1,BA}. First of all, 
if we generalize model {\it F\,} by replacing the complex field $\psi$ by a 
vector $\Psi=(\psi_1,\ldots,\psi_n)$ of $n$ complex fields, then it turns out 
that the Hartree approximation is exact for the Green's function $G$ in the 
limit $n\rightarrow\infty$. For each closed loop of thick solid lines there 
will be a factor $n$. If we rescale the coupling parameters according to 
$\tilde u_0\sim n^{-1}$, $\gamma_0\sim n^{-1/2}$, and $g_0\sim n^{-1/2}$ so 
that $B_{3,\alpha\beta\gamma}\sim n^{-1/2}$ and 
$B_{4,\alpha\beta,\gamma\delta}\sim n^{-1}$, then only the tadpole diagrams 
shown in Fig.\ \ref{F02}b will be nonzero in the limit $n\rightarrow\infty$. 
It is well known that models involving Ginzburg-Landau functionals like the
$\phi^4$ model can be solved exactly in this limit (see 
e.g.\ Ref.\ \onlinecite{A1}). The same is true also for model {\it F}. 
While in the limit $n\rightarrow\infty$ the RG theory is not needed because
the Hartree approximation is exact, in our case for $n=1$ the RG theory is
necessary to obtain the correct critical behavior of the physical quantities
near $T_\lambda$.

We may expect that beyond the Hartree approximation an $1/n$ expansion may 
yield the proper corrections. However, this is not true. By experience with
quantum-field theory of many-particle systems with degeneracies \cite{H2,HB}
we have found that proper corrections are given by the modified 
self-consistent random-phase approximation (modified SC-RPA), where the 
modification is a gauge transformation which implies bosonization for a
proper treatment of the low energetic collective excitations. The method has
been invented for two-dimensional electron systems in the regime of the 
fractional quantum Hall effect \cite{H2} and tested for simple exactly 
solvable models \cite{HB}. We have found that the modified SC-RPA can describe
the superfluid transition and Bose-Einstein condensation in interacting 
boson systems where the average order parameter is $\langle\psi\rangle=0$ due
to phase fluctuations above and below the transition \cite{HB}. For this
reason we believe that the modified SC-RPA combined with the RG theory will 
be successful in the present case for model {\it F}. 

The main feature of the modified SC-RPA is that it implies a nontrivial
spectrum for the Green's function $G$ while the Hartree approximation does 
not. This fact is important in quantum many-particle physics because the
spectra of the quasiparticles are important physical results. However, in the
present case for critical phenomena and second-order phase transitions 
nontrivial spectra are not essential while the application of the RG theory
is important. For this reason, the Hartree approximation combined with the 
RG theory should be sufficient for our purposes. Even though the modified 
SC-RPA combined with the RG theory would be desirable, we expect only some 
corrections while the calculations would be much more complicated. 

\subsection{Evaluation of the Green's function}
Now, we evaluate the self energy $\Sigma$ and determine the unrenormalized
Green's function $G$ in Hartree approximation. The tadpole diagrams in 
Fig.\ \ref{F02}b imply a self energy of the form
\begin{equation}
\Sigma_{\alpha\beta}({\bf r},t;{\bf r}^\prime,t^\prime) = 
- \Delta K_{\alpha\beta} \, \delta({\bf r} - {\bf r}^\prime) 
\, \delta(t-t^\prime) 
\label{B03}
\end{equation}
where $\Delta K_{\alpha\beta}$ are the elements of a $2\times 2$ matrix 
which depend on the space coordinate $z$ but do not contain differential
operators. Applying the Feynman rules of Sec.\ \ref{S02} we obtain the matrix
\begin{equation}
(\Delta K_{\alpha\beta}) = \pmatrix{ 0 & {1\over 2} \Delta L \cr 
{1\over 2} \Delta L^+ & 0 \cr } 
\label{B04}
\end{equation}
where
\begin{equation}
\Delta L = 2 b_3 N_0^{-1} [ \lambda_0 \gamma_0 \bbox{\nabla}^2 n_{\rm s} + 
g_0 \bbox{\nabla} {\bf J}_{\rm s} ]  +  2 b_4 n_{\rm s}  \ .
\label{B05}
\end{equation}
Here
\begin{equation}
n_{\rm s} = \langle \vert\psi\vert^2 \rangle = G_{22} ({\bf r},t;{\bf r},t) 
\label{B06}
\end{equation}
is a density related to the entropy and 
\begin{equation}
{\bf J}_{\rm s} = \langle {\rm Im} [ \psi^* \bbox{\nabla} \psi ] \rangle = 
\lim_{{\bf r}^\prime \to {\bf r}} {\rm Im} [ \bbox{\nabla} G_{22} 
({\bf r},t;{\bf r}^\prime,t) ] 
\label{B07}
\end{equation}
is the superfluid current density. $N_0^{-1}$ is the inverse of the 
differential operator (\ref{A10}). Since $n_{\rm s}$ and ${\bf J}_{\rm s}$ 
in (\ref{B05}) depend only on $z$, we may drop the time derivative in 
$N_0^{-1}$. Then, from (\ref{B05}) we obtain
\begin{equation}
\Delta L = 2 (b_4 - \chi_0 \gamma_0 b_3) n_{\rm s} - 2 b_3 (\chi_0 g_0 /
\lambda_0) (\bbox{\nabla}^2)^{-1} (\bbox{\nabla} {\bf J}_{\rm s}) \ . 
\label{B08}
\end{equation}
Inserting $b_3$ and $b_4$ of (\ref{A13}) we obtain
\begin{equation}
b_4 - \chi_0 \gamma_0 b_3 = 2\Gamma_0 (\tilde u_0 - \textstyle{1\over 2}
\chi_0 \gamma_0^2 ) = 2\Gamma_0 u_0 \ ,
\label{B09}
\end{equation}
where $u_0=\tilde u_0 - {1\over 2}\chi_0\gamma_0^2$ is the effective coupling 
between the $\psi$ fields in thermal equilibrium after the entropy field $m$
has been integrated out (see Ref.\ \onlinecite{D1}). Since the heat current 
$Q$ flows in $z$ direction, only the $z$ component $J_{{\rm s,}z}=J_{\rm s}$ 
of the superfluid current is nonzero so that
\begin{equation}
\label{B10}
(\bbox{\nabla}^2)^{-1} (\bbox{\nabla} {\bf J}_{\rm s}) = \partial_z^{-1}
J_{\rm s} \ , 
\end{equation}
Thus, from (\ref{B08}) we obtain
\begin{eqnarray}
\Delta L &=& \Gamma_0 [ 4u_0 n_{\rm s} - 2 \chi_0 \gamma_0 (g_0/\lambda_0)
\,\partial_z^{-1} J_{\rm s} ] \nonumber\\
&& \hspace{1cm} + i g_0 (g_0/\lambda_0) \,\partial_z^{-1} J_{\rm s} \ . 
\label{B11}
\end{eqnarray}
Now, we define
\begin{equation}
L = L_0 + \Delta L \ .
\label{B12}
\end{equation}
Then from (\ref{A09}) and (\ref{B11}) we obtain
\begin{equation}
L = \partial_t + \Gamma_0 [ r_1(z) - \bbox{\nabla}^2 ] - i {g_0 \over 2\chi_0
\gamma_0} \Delta r_0(z) 
\label{B13}
\end{equation}
where 
\begin{eqnarray}
r_1(z) &=& \tau_0(z) + 2\chi_0 \gamma_0 \Bigl[ {1\over\chi_0} m_{\rm mf}(z)
- {g_0\over\lambda_0} \,\partial_z^{-1} J_{\rm s} \Bigr] \nonumber\\
&&+ 4 u_0 n_{\rm s} 
\label{B14}
\end{eqnarray}
and
\begin{equation}
\Delta r_0(z) = 2\chi_0 \gamma_0 \Bigl[ {1\over\chi_0} m_{\rm mf}(z) -h_0
- {g_0\over\lambda_0} \,\partial_z^{-1} J_{\rm s} \Bigr] 
\label{B15}
\end{equation}
are effective parameters. Furthermore we define the matrix
\begin{equation}
K_{\alpha\beta} = K_{0,\alpha\beta} + \Delta K_{\alpha\beta} 
\label{B16}
\end{equation}
and obtain
\begin{equation}
(K_{\alpha\beta}) = \pmatrix{ -\Gamma_0^\prime & {1\over 2} L \cr
{1\over 2} L^+ & 0 \cr}  \ . 
\label{B17}
\end{equation}
Now, the Green's function $G$ in Hartree approximation is obtained easily.
We find that (\ref{B12}) and (\ref{B16}) are equivalent to the Dyson 
equation (\ref{B02}). Thus, as a result we obtain
\begin{equation}
G_{\alpha\beta}({\bf r},t;{\bf r}^\prime,t^\prime) = K^{-1}_{\alpha\beta} 
\, \delta({\bf r} - {\bf r}^\prime) \, \delta(t-t^\prime) 
\label{B18}
\end{equation}
where $K_{\alpha\beta}$ is given by (\ref{B17}) together with (\ref{B13}). 
Clearly, the Green's function $G$ in Hartree approximation has the same 
structure as the free Green's function $G_0$ where just the parameters have
been replaced by effective parameters. As a consequence, the following 
calculations are considerably simplified because we may restrict the
considerations to the effective parameters $r_1(z)$ and $\Delta r_0(z)$ only.
We will derive self-consistent equations for $r_1(z)$ and $\Delta r_0(z)$ to
determine the effective parameters. Eventually, the Green's function $G$ is
obtained from (\ref{B18}) together with (\ref{B13}) and (\ref{B17}). 

\subsection{Evaluation of $n_{\rm s}$ and $J_{\rm s}$}
Next we evaluate $n_{\rm s}$ and $J_{\rm s}$ by inserting the Green's 
function (\ref{B18}) into (\ref{B06}) and (\ref{B07}). In 
Ref.\ \onlinecite{HD1} $n_{\rm s}$ and $J_{\rm s}$ were evaluated for the 
free Green's function so that here we need not perform these calculations 
once again. Since the Green's function in Hartree approximation has the same 
structure, we may use the results of Ref.\ \onlinecite{HD1} with a slight
modification. We just need to replace the free parameters by the effective 
parameters $r_1(z)$ and $\Delta r_0(z)$ appropriately. The basic assumption 
of the previous calculations \cite{HD1} was that the parameters $r_1(z)$ and 
$\Delta r_0(z)$ in the Green's function are linear functions of $z$. For the 
free Green's function, where $n_{\rm s}$ and $J_{\rm s}$ in (\ref{B14}) and 
(\ref{B15}) are omitted, this is indeed true, because $m_{\rm mf}(z)$ defined 
in (\ref{A06}) and $\tau_0(z)$, which is related to $T_\lambda(z)$, are 
linear functions of $z$. Thus, in the present case we must assume as an 
approximation, that $r_1(z)$ and $\Delta r_0(z)$ are linearized locally so 
that only the slopes $r_1^\prime$ and $\Delta r_0^\prime$ are included but 
the curvatures and the higher-order derivatives are neglected. We will later 
show in Sec.\ \ref{S07} that this assumption is justified. 

Now, from (3.22) and (3.25) in the second paper of Ref.\ \onlinecite{HD1} we 
obtain
\begin{equation}
n_{\rm s} = 2\ \Phi_{-1+\epsilon/2}(X) \  \int {d^dp\over (2\pi)^d} \,
{1\over r_1(z) + {\bf p}^2} \ , 
\label{B19}
\end{equation}
and 
\begin{equation}
J_{\rm s} = {g_0\over 2\Gamma_0^\prime} \, {\Delta r_0^\prime\over 2 \chi_0
\gamma_0} \ \Phi_{\epsilon/2}(X) \  \int {d^dp\over (2\pi)^d} \,
{1\over [ r_1(z) + {\bf p}^2 ]^2 } \ . 
\label{B20}
\end{equation}
Here $\epsilon$ is related to the space dimension $d$ by $\epsilon = 4-d$. 
The function $\Phi_\alpha(X)$ is defined by the asymptotic series
\begin{equation}
\Phi_\alpha(X) = \sum_{N=0}^\infty {\Gamma(\alpha + 3N) \over \Gamma(\alpha)}
\ {X^N\over N!} 
\label{B21}
\end{equation}
and contains all the effects beyond linear response theory. In the integrals
the parameter $\tilde r_0(z)$ of Ref.\ \onlinecite{HD1} has been replaced by 
the effective parameter $r_1(z)$. The dimensionless parameter $X$ is given 
here by
\begin{eqnarray}
X &=& {1\over 12 [r_1(z)]^3} \Bigl[ r_1^{\prime2} + 2 {\Gamma_0^{\prime\prime}
\over \Gamma_0^\prime} \Bigl( {g_0 \over 4\chi_0 \gamma_0 \Gamma_0^\prime}
\Delta r_0^\prime \Bigr) r_1^\prime \nonumber\\
&& \hspace{2cm} - \Bigl( {g_0 \over 4\chi_0 \gamma_0 \Gamma_0^\prime}
\Delta r_0^\prime \Bigr)^2 \Bigr]  
\label{B22}
\end{eqnarray}
where
\begin{equation}
r_1^\prime = \partial_z r_1(z) \ , \hspace{1cm} \Delta r_0^\prime = 
\partial_z \Delta r_0(z) \ . 
\label{B23}
\end{equation}
The entropy current $q$ in the formulas of Ref.\ \onlinecite{HD1} is here
replaced by $r_1^\prime$ or $\Delta r_0^\prime$ times $-\lambda_0/2\chi_0
\gamma_0$. We note that our formulas reduce to those of Ref.\ \onlinecite{HD1}
if we omit $n_{\rm s}$ and $J_{\rm s}$ in the effective parameters 
(\ref{B14}) and (\ref{B15}) and if we neglect gravity. The integrals can be 
evaluated \cite{HD1,D1} in dimensional regularization so that we obtain
\begin{equation}
n_{\rm s}(z) = - {2\over \epsilon} A_d\ \Phi_{-1+\epsilon/2}(X) \ 
[r_1(z)]^{1-\epsilon/2} \ , 
\label{B24}
\end{equation}
and 
\begin{equation}
J_{\rm s}(z) = {g_0\over 2\Gamma_0^\prime} \, {\Delta r_0^\prime\over 2\chi_0
\gamma_0} \, {1\over \epsilon} A_d \ \Bigl(1- {\epsilon\over 2} \Bigr) 
\ \Phi_{\epsilon/2}(X) \, [r_1(z)]^{-\epsilon/2} 
\label{B25}
\end{equation}
where $A_d=S_d \Gamma(1-\epsilon/2) \Gamma(1+\epsilon/2)$, $S_d = \Omega_d
/(2\pi)^d$, and $\Omega_d = 2\pi^{d/2} / \Gamma(d/2)$ is the surface of the
$d$ dimensional unit sphere. Clearly, $n_{\rm s}(z)$ and $J_{\rm s}(z)$ 
depend on $z$ implicitly via $r_1(z)$ and the derivatives $r_1^\prime$ and
$\Delta r_0^\prime$. 

\subsection{Self-consistent equations for \break the effective parameters}
Eqs.\ (\ref{B14}) and (\ref{B15}) together with (\ref{B24}) and (\ref{B25})
are self-consistent equations for the effective parameters $r_1(z)$ and
$\Delta r_0(z)$. The structure of these equations can be simplified. First
of all we note that the effective parameters and its derivatives can be 
related to the temperatures $T(z)$, $T_\lambda(z)$, and the heat current $Q$.
We find 
\begin{equation}
\Delta r_0(z) = 2\chi_0 \gamma_0 \Bigl\langle {\delta H \over \delta m} \Bigr
\rangle = 2 \chi_0 \gamma_0 \, {T(z) - T_0 \over T_\lambda} \ . 
\label{B26}
\end{equation}
To prove this relation $\langle \delta H/ \delta m\rangle$ can be evaluated
explicitly in Hartree approximation and compared with (\ref{B15}). 
Equivalently, we take the average of (\ref{A02}) and obtain $\partial_t 
\langle m \rangle + \bbox{\nabla} {\bf q} = 0$ where ${\bf q}$ 
\begin{equation}
{\bf q} = -\lambda_0 \bbox{\nabla} \Bigl\langle {\delta H\over \delta m}
\Bigr\rangle - g_0 {\bf J}_{\rm s} 
\label{B27}
\end{equation}
is the entropy current. Since $z$ is the only space coordinate, this equation
can be rewritten in the form 
\begin{eqnarray}
{\partial\over\partial z} \Bigl\langle {\delta H\over \delta m} \Bigr\rangle 
&=& - {q\over \lambda_0} - {g_0 \over \lambda_0} J_{\rm s} \nonumber\\
&=& {\partial\over\partial z} \Bigl[ {1\over\chi_0} m_{\rm mf}(z) - h_0
- {g_0\over \lambda_0} \,\partial_z^{-1} J_{\rm s} \Bigr] \ . 
\label{B28}
\end{eqnarray}
Thus, integrating this equation and comparing with (\ref{B15}) we obtain 
(\ref{B26}). We note that (\ref{B15}) contains integration constants via 
$m_{\rm mf}(z)$ and $\partial_z^{-1}J_{\rm s}$. Since $m$ is the entropy
density divided by $k_{\rm B}$, the quantity $\langle \delta H/ \delta m
\rangle$ is a temperature difference divided by $T_\lambda$. This fact 
explains the last equality sign in (\ref{B26}). Here $T_0$ is a constant
reference temperature which may be arbitrary. In the denominator the $z$ 
dependence of $T_\lambda$ due to gravity is very small and may be neglected. 

The temperature parameter $r_0(z)$ is defined by \cite{HD1}
\begin{eqnarray}
r_0(z) &=& \tau_0(z) + 2\chi_0\gamma_0 \Bigl( h_0 + \Bigl\langle 
{\delta H\over \delta m} \Bigr\rangle \Bigr) \nonumber\\ 
&=& \tau_0(z) + 2\chi_0 \gamma_0 h_0 + \Delta r_0(z) \ .
\label{B29}
\end{eqnarray}
In thermal equilibrium this parameter is the coefficient of $\vert\psi\vert^2$
in the energy functional after the entropy variable $m$ has been integrated
out \cite{D1}. It is related to the temperature by
\begin{equation}
r_0(z) - r_{\rm 0c} = 2\chi_0\gamma_0 \, {T(z)-T_\lambda(z)\over T_\lambda}
= 2\chi_0\gamma_0 \, {\Delta T(z)\over T_\lambda} \ . 
\label{B30}
\end{equation}
The critical value is $r_{\rm 0c}=0$ in one-loop approximation \cite{D1} 
and hence also in Hartree approximation. We find that the first line on the
right-hand side of (\ref{B14}) is identified by $r_0(z)$. Thus, 
Eq.\ (\ref{B14}) simplifies into 
\begin{equation}
r_1(z) = r_0(z) + 4 u_0 n_{\rm s} \ . 
\label{B31}
\end{equation}
We resolve this equation with respect to $r_0(z)$ and insert (\ref{B24}) for
$n_{\rm s}$. Then we obtain 
\begin{equation}
r_0(z) = r_1(z) \Bigl\{ 1 + {8 u_0 \over \epsilon}\, A_d \, 
\Phi_{-1+\epsilon/2}(X) \, [r_1(z)]^{-\epsilon/2} \Bigr\} \ . 
\label{B32}
\end{equation}
Next, we take the derivative of this equation with respect to $z$ and obtain
\begin{equation}
r_0^\prime = r_1^\prime \Bigl\{ 1 + {8 u_0 \over \epsilon}\, A_d \, 
\Bigl(1 - {\epsilon\over 2} \Bigr) \, \Phi_{\epsilon/2}(X) \, 
[r_1(z)]^{-\epsilon/2} \Bigr\} \ . 
\label{B33}
\end{equation}
Furthermore, the derivative of (\ref{B15}) yields 
\begin{equation}
\Delta r_0^\prime = -2\chi_0\gamma_0 \Bigl[ {q\over \lambda_0} + {g_0\over
\lambda_0}\, J_{\rm s} \Bigr] \ . 
\label{B34}
\end{equation}
Resolving this equation with respect to $q$ and inserting (\ref{B25}) for
$J_{\rm s}$ we obtain 
\begin{eqnarray}
q = {Q\over k_{\rm B} T_\lambda} &=& - \lambda_0 \Bigl\{ 1 
+ {g_0^2 \over 2\lambda_0 \Gamma_0^\prime}\, {1 \over \epsilon}\, A_d \, 
\Bigl( 1 - {\epsilon\over 2} \Bigr)  \nonumber\\ 
&&\times \Phi_{\epsilon/2}(X) \, [r_1(z)]^{-\epsilon/2} \Bigr\} 
{\Delta r_0^\prime \over 2\chi_0\gamma_0} \ . 
\label{B35}
\end{eqnarray}
On the other hand from (\ref{B26}) we find that $\Delta r_0^\prime$ is 
related to the temperature gradient $\partial_z T$ by 
\begin{equation}
\Delta r_0^\prime = 2\chi_0 \gamma_0 T_\lambda^{-1} \partial_z T \ . 
\label{B36}
\end{equation}
Finally, Eq.\ (\ref{B29}) implies
\begin{equation}
r_0^\prime = \Delta r_0^\prime - 2\chi_0\gamma_0 T_\lambda^{-1} \partial_z
T_\lambda \ , 
\label{B37}
\end{equation}
where we have identified $\tau_0^\prime = -2\chi_0\gamma_0 T_\lambda^{-1}
\partial_z T_\lambda$. Clearly, the difference between $r_0^\prime$ and
$\Delta r_0^\prime$ is due to the gradient of $T_\lambda(z)$ which is the 
effect of gravity. Thus, in a microgravity environment $r_0^\prime$ and
$\Delta r_0^\prime$ are equal. 

Now, the self-consistent equations which allow the determination of all the
effective parameters are given by (\ref{B32}), (\ref{B33}), and (\ref{B35})
together with (\ref{B26}), (\ref{B30}), (\ref{B37}), and (\ref{B22}). These
are seven equations for seven unknown variables $r_0(z)$, $\Delta r_0(z)$,
$r_1(z)$, $r_0^\prime$, $\Delta r_0^\prime$, $r_1^\prime$, and $X$. As an
input we need the temperatures $T(z)$, $T_\lambda(z)$, the difference
$\Delta T(z)=T(z)-T_\lambda(z)$, the heat current $Q=k_{\rm B}T_\lambda q$, 
and the gradient $\partial_z T_\lambda$ for a given space variable $z$. 
Since the seven equations do not depend explicitly on $z$, we do not need 
the temperature profiles as functions of $z$. Instead, we obtain the 
temperature gradient $\partial_z T$ from (\ref{B36}) so that the temperature
profile $T(z)$ can be calculated by integration. Eventually, for given 
$\Delta T$ and $Q$ physical quantities like the specific heat and the thermal
conductivity can be calculated. This will be done in Secs.\ \ref{S05} and 
\ref{S08}.

\section{Renormalization and application of the renormalization-group
theory} \label{S04}
In (\ref{B32}), (\ref{B33}), and (\ref{B35}) the first-order terms exhibit 
infrared divergences at criticality where $r_1(z)\rightarrow 0$ while the
function $\Phi_\alpha(X)$ is of order unity. For this reason the 
renormalization of these equations and the application of the RG theory are
necessary to achieve a resummation of the infrared divergences and a proper 
treatment of the critical fluctuations. We use the concept of renormalization 
by minimal subtraction of dimensional poles. The calculations are performed
at fixed dimension $d=4-\epsilon$ (i.e.\ no $\epsilon$ expansion is applied).
For model {\it F\,} this renormalization scheme is described in 
Ref.\ \onlinecite{D1}. The renormalization factors of the fields $\psi$ and 
$\tilde\psi$ are $Z_\psi=Z_{\tilde\psi}=1$ in one-loop order and hence also 
in Hartree approximation. Thus, the Green's function $G$ is not renormalized 
here. As a consequence, also the operator $L$ defined in (\ref{B13}) is not
renormalized. The parameter $\Gamma_0$ is renormalized according to \cite{D1}
$\Gamma_0 = Z_\Gamma^{-1} \Gamma$ where, however, in one-loop order and in
Hartree approximation $Z_\Gamma=1$. Thus, from (\ref{B13}) we conclude that 
$r_1(z)$ is not renormalized. 

The parameters $\gamma_0$ and $g_0$ are renormalized according to \cite{D1}
\begin{eqnarray}
\chi_0\gamma_0 &=& \gamma\, (\chi_0 Z_m)^{1/2}Z_r (\mu^\epsilon/A_d)^{1/2} \ , 
\label{C01}  \\
g_0 &=& g\, (\chi_0 Z_m)^{1/2} (\mu^\epsilon/A_d)^{1/2} \ . 
\label{C02}
\end{eqnarray}
Furthermore, we renormalize $\Delta r_0(z)= Z_r \Delta r(z)$. Consequently,
we find 
\begin{equation}
{g_0 \over 2\chi_0 \gamma_0}\, \Delta r_0(z) = {g\over 2\gamma}\, \Delta r(z)
\label{C03}
\end{equation}
which is consistent with the requirement that the last term in (\ref{B13}) 
is not renormalized. Thus, in terms of the renormalized parameters the 
operator (\ref{B13}) reads
\begin{equation}
L = \partial_t + \Gamma [ r_1(z) - \bbox{\nabla}^2 ] - i (g/2\gamma) 
\Delta r(z) \ .
\label{C04}
\end{equation}
Next we renormalize (\ref{B32}). For this purpose we need the relations 
\cite{D1} 
\begin{eqnarray}
r_0(z) - r_{\rm 0c} &=& Z_r r(z) \ , 
\label{C05}  \\
u_0 &=& u\, Z_u Z_\psi^{-2} (\mu^\epsilon/A_d) 
\label{C06}
\end{eqnarray}
where $r_{\rm 0c}=0$ and $Z_\psi=1$ in Hartree approximation. We separate the
ultraviolet divergence on the right-hand side of (\ref{B32}) which here in
dimensional regularization is a pole $\sim 1/\epsilon$. By choosing
\begin{equation}
Z_r = Z_u = 1/ [1 - 8u/\epsilon] 
\label{C07}
\end{equation}
the ultraviolet divergence is canceled. Eventually we obtain
\begin{equation}
r(z) = r_1(z) \Bigl\{ 1 + {8u\over\epsilon}  \Bigl[ \Phi_{-1+\epsilon/2}(X) 
\, \Bigl({r_1(z)\over\mu^2}\Bigr)^{-\epsilon/2} - 1 \Bigr] \Bigr\} \ . 
\label{C08}
\end{equation}
Analogously we separate the ultraviolet divergence on the right-hand side of
(\ref{B33}). Using (\ref{C05})-(\ref{C07}) we obtain
\begin{equation}
r^\prime = r_1^\prime \Bigl\{ 1 + {8u\over\epsilon}  \Bigl[ \Bigl( 1 - 
{\epsilon\over 2} \Bigr)\, \Phi_{\epsilon/2}(X) \, \Bigl({r_1(z)\over\mu^2}
\Bigr)^{-\epsilon/2} - 1 \Bigr] \Bigr\} \ . 
\label{C09}
\end{equation}
We replace the parameters in the dimensionless variable $X$ defined in 
(\ref{B22}) by the renormalized ones. It turns out that all $Z$ factors 
cancel so that $X$ is not renormalized. Thus, we obtain
\begin{eqnarray}
X &=& {1\over 12 [r_1(z)]^3} \nonumber\\
&&\times \Bigl[ r_1^{\prime2} + 2 {\Gamma^{\prime\prime}
\over \Gamma^\prime} \Bigl( {g \over 4\gamma \Gamma^\prime} \Delta r^\prime 
\Bigr) r_1^\prime - \Bigl( {g \over 4\gamma \Gamma^\prime} \Delta r^\prime 
\Bigr)^2 \Bigr] \ .  
\label{C10}
\end{eqnarray}
For convenience we replace the renormalized couplings by the dimensionless 
combinations \cite{D1} $w=\Gamma/\lambda$, $F=g/\lambda$, and $f=F^2/
w^\prime$. Furthermore, we introduce the dimensionless effective parameters
\begin{eqnarray}
\rho &=& r(z)/\mu^2 \ , \hspace{1.5cm} \rho^\prime = r^\prime/\mu^3 \ , 
\label{C11}  \\
\Delta\rho &=& \Delta r(z)/\mu^2 \ , \hspace{0.9cm} \Delta\rho^\prime = 
\Delta r^\prime/\mu^3 \ , \label{C12}  \\
\rho_1 &=& r_1(z)/\mu^2 \ , \hspace{1.3cm} \rho_1^\prime = 
r_1^\prime/\mu^3 \ . \label{C13}
\end{eqnarray}
Since the coordinate $z$ does not appear explicitly in the self-consistent
equations for the effective parameters, we omit $z$ as an argument from
now on. Then, Eqs.\ (\ref{C08})-(\ref{C10}) can be rewritten as
\begin{eqnarray}
\rho &=& \rho_1 \Bigl\{ 1 + {8u\over\epsilon} \Bigl[ \Phi_{-1+\epsilon/2}(X) 
\, \rho_1^{-\epsilon/2} - 1 \Bigr] \Bigr\} \ ,  \label{C14}  \\
\rho^\prime &=& \rho_1^\prime \Bigl\{ 1 + {8u\over\epsilon} \Bigl[ \Bigl( 
1 - {\epsilon\over 2} \Bigr)\, \Phi_{\epsilon/2}(X) \, \rho_1^{-\epsilon/2} 
- 1 \Bigr] \Bigr\} \ ,  \label{C15}  \\
X &=& {1\over 12 \rho_1^3} \Bigl[ \rho_1^{\prime2} + 2 {w^{\prime\prime} 
\over w^\prime} \Bigl( {F \over 4\gamma w^\prime} \Delta \rho^\prime \Bigr) 
\rho_1^\prime - \Bigl( {F \over 4\gamma w^\prime} \Delta \rho^\prime 
\Bigr)^2 \Bigr] \ ,  \nonumber\\ \label{C16}
\end{eqnarray}
respectively. 

The entropy current is renormalized by \cite{HD1} 
\begin{equation}
q=(\chi_0 Z_m)^{1/2} q^{\rm ren} \ . 
\label{C17}
\end{equation}
This equation together with (\ref{C02}) implies that the ratio
\begin{equation}
Q/g_0k_{\rm B} T_\lambda = q/g_0 = (q^{\rm ren}/g)\, (A_d/\mu^\epsilon)^{1/2}
\label{C18}
\end{equation}
need not be renormalized because the $Z$ factors cancel. In (\ref{B35}) we 
separate the ultraviolet divergence and replace the coupling parameters by the 
renormalized parameters. Additionally, we need the renormalization \cite{D1}
$\lambda_0 = \chi_0 Z_\lambda^{-1} \lambda$. Using the $Z$ factor product
\begin{equation}
Z_m Z_\lambda = 1/[1-f/2\epsilon] 
\label{C19}
\end{equation}
we find that the ultraviolet divergence is canceled. Then, in terms of the 
renormalized couplings and effective parameters we obtain
\begin{eqnarray}
{Q \, \mu^{\epsilon-3} \over g_0 k_{\rm B} T_\lambda} &=& 
- {A_d \over 2\gamma F} \Bigl\{ 1 + {f \over 2\epsilon}  \nonumber\\ 
&&\times \Bigl[ \Bigl( 1 - {\epsilon\over 2} \Bigr)  
\, \Phi_{\epsilon/2}(X) \, \rho_1^{-\epsilon/2} - 1 \Bigr] \Bigr\} 
\Delta \rho^\prime 
\label{C20}
\end{eqnarray}
where the left-hand side need not be renormalized because of (\ref{C18}). 
Finally, we renormalize (\ref{B26}), (\ref{B30}), (\ref{B36}), and 
(\ref{B37}) and obtain
\begin{eqnarray}
\Delta \rho &=& \tau^{-1} [T(z) - T_0]/T_\lambda \ ,  \label{C21}  \\
\rho &=& \tau^{-1} [T(z) - T_\lambda(z)]/T_\lambda \ ,  \label{C22}  \\
\Delta \rho^\prime &=& \tau^{-1} (\mu T_\lambda)^{-1} \partial_z T  \ , 
\label{C23}  \\
\rho^\prime &=& \Delta\rho^\prime - \tau^{-1} (\mu T_\lambda)^{-1} 
\partial_z T_\lambda  \ ,
\label{C24} 
\end{eqnarray}
respectively, where 
\begin{equation}
\tau = \Bigl( {A_d \mu^d \over\chi_0 Z_m} \Bigr)^{1/2} \,{1\over 2\gamma} \ . 
\label{C25}
\end{equation}
In these equations $Z_m$ does not cancel. We note that the renormalization is
exact in all the above equations. This means that we need not expand the $Z$ 
factors in powers of the renormalized couplings. The reason of this fact is 
that the Hartree approximation is exact in the limit $n\rightarrow\infty$ for 
model {\it F\,} with an $n$-component complex order parameter. On the other 
hand, the $Z$ factors do not agree with those of the previous theories 
\cite{D1} in one loop-order because the Hartree approximation is not a loop 
expansion. (The correct one-loop $Z$ factors would be obtained if we would 
consider a Hartree-Fock approximation and include both the Hartree and the 
Fock term in the self energy $\Sigma$.) 

By the renormalization a characteristic length scale is introduced which is
described by the parameter $\mu$. The RG theory is based on the fact that 
this length scale is arbitrary and may be changed according to 
$\mu\rightarrow \mu l$, where $l$ is the RG flow parameter. As a consequence, 
the renormalized coupling parameters $u(l)$, $\gamma(l)$, $w(l)$, $F(l)$, and 
$f(l)$ depend on $l$. Furthermore, also the $Z$ factors depend on $l$. Now, 
the dimensionless parameter defined in 
(\ref{C25}) reads
\begin{equation}
\tau = \Bigl( {A_d (\mu l)^d \over\chi_0 Z_m(l)} \Bigr)^{1/2} 
\,{1\over 2\gamma(l)} \ . 
\label{C26}
\end{equation}
For convenience we will use $\tau$ as the RG flow parameter instead of 
$l$ because $\tau$ is closely related to the reduced temperature by 
(\ref{C22}) and the renormalized coupling parameters $u[\tau]$, 
$\gamma[\tau]$, $w[\tau]$, $F[\tau]$, and $f[\tau]$ were determined as
functions of $\tau$ in Ref.\ \onlinecite{D1}. We identify $\mu l = \xi^{-1}$
by the correlation length $\xi=\xi(\tau)$, which in the asymptotic region
is given by $\xi(\tau)=\xi_0 \tau^{-\nu}$. The identification $\mu l = 
\xi^{-1}$ is correct in one-loop order, corrections appear in higher orders 
\cite{SD}.

Now, we write the self-consistent equations for the effective parameters in
a form which is appropriate for the numerical evaluation. For this purpose
we eliminate some of the dimensionless parameters and introduce some new 
parameters. First of all we note that the asymptotic series (\ref{B21}) is not
useful to evaluate the function $\Phi_\alpha(X)$. In Ref.\ \onlinecite{HD1} an
integral representation was found by 
\begin{equation}
\Phi_\alpha(X) = [\Gamma(\alpha)]^{-1} \zeta^\alpha {\cal F}_\alpha(\zeta) 
\label{C27}
\end{equation}
where $\zeta = (-X)^{-1/3}$ and 
\begin{equation}
{\cal F}_\alpha(\zeta) =\int_0^\infty dv\, v^{\alpha-1} \exp(-v^3-v\zeta) \ .
\label{C28}
\end{equation}
The integral is well defined for $\alpha>0$. For $\alpha<0$ the function 
${\cal F}_\alpha(\zeta)$ is obtained by analytical continuation in $\alpha$ 
or equivalently by partial integration in (\ref{C28}) to remove the 
ultraviolet divergence. Now, we introduce the parameter
\begin{equation}
\sigma = - {1\over 12} \Bigl[ \rho_1^{\prime2} + 2 {w^{\prime\prime}[\tau] 
\over w^\prime[\tau]} \Bigl( {F[\tau] \, \Delta \rho^\prime \over 
4\gamma[\tau] w^\prime[\tau]} \Bigr) \rho_1^\prime - \Bigl( {F[\tau] 
\, \Delta \rho^\prime \over 4\gamma[\tau] w^\prime[\tau]} \Bigr)^2 \Bigr] 
\label{C29}
\end{equation}
so that $X=-\sigma/\rho_1^3$ or equivalently $\rho_1 = \sigma^{1/3} \zeta$. 
In the following we eliminate $\rho_1$ and $X$ in favor of $\zeta$ and 
$\sigma$. For convenience we define the amplitudes
\begin{eqnarray}
A &=& \epsilon^{-1} \bigl[ \Phi_{-1+\epsilon/2}(X) \, \rho_1^{-\epsilon/2} 
- 1 \bigr]  \label{C30}  \\
&=& {1\over\epsilon} \Bigl[{\sigma^{-\epsilon/6} \over \Gamma(-1+\epsilon/2)} 
\,\zeta^{-1} \,{\cal F}_{-1+\epsilon/2}(\zeta) - 1 \Bigr] \ , \label{C31}  \\ 
A_1 &=& \epsilon^{-1} \bigl[ (1-\epsilon/2) \, \Phi_{\epsilon/2}(X) \, 
\rho_1^{-\epsilon/2} - 1 \bigr]  \label{C32}  \\
&=& {1\over\epsilon} \Bigl[ - {\sigma^{-\epsilon/6} \over 
\Gamma(-1+\epsilon/2)} \, {\cal F}_{\epsilon/2}(\zeta) - 1 \Bigr] \ .  
\label{C33} 
\end{eqnarray}
Then we rewrite (\ref{C14}) as
\begin{equation}
\rho = \sigma^{1/3} \zeta \, \bigl\{ 1 + 8u[\tau] A \bigl\} \ .
\label{C34}
\end{equation}
To eliminate $\rho$ we insert this into (\ref{C22}). Resolving with respect 
to the temperature difference we obtain
\begin{equation}
\Delta T(z) = T(z) - T_\lambda(z) = T_\lambda \, \tau \, \sigma^{1/3} \zeta
\,\bigl\{ 1 + 8u[\tau] A \bigr\} \ . 
\label{C35}
\end{equation}
Next we resolve (\ref{C20}) with respect to $\Delta \rho^\prime$ and obtain
\begin{equation}
\Delta \rho^\prime = - {2\gamma[\tau] F[\tau]\over A_d} \, \Bigl( {Q\xi^{d-1}
\over g_0 k_{\rm B} T_\lambda} \Bigr) \Big/ \bigl\{ 1 + (f[\tau]/2) A_1 
\bigr\} \ . 
\label{C36}
\end{equation}
Resolving (\ref{C15}) with respect to $\rho_1^\prime$ and eliminating 
$\rho^\prime$ by inserting (\ref{C24}) we obtain 
\begin{equation}
\rho_1^\prime = \Bigl[ \Delta\rho^\prime - {1\over\tau} {\xi\over T_\lambda}
\, {\partial T_\lambda\over \partial z} \Bigr] \Big/ \bigl\{ 1 + 
8u[\tau] A_1 \bigr\} \ . 
\label{C37}
\end{equation}
Finally, from (\ref{C23}) we obtain the temperature gradient
\begin{equation}
\partial_z T = T_\lambda \, (\tau/\xi) \, \Delta \rho^\prime  \ . 
\label{C38}
\end{equation}

Until now the RG flow parameter $\tau$ is arbitrary. We must choose $\tau$
so that an optimum resummation of the infrared divergences in the 
perturbation series is achieved. From our experience we find that the 
condition
\begin{equation}
\sigma^{1/3}\, ( 8 + \zeta - 16 u[\tau] A\, \zeta ) = 1 
\label{C39}
\end{equation}
is an optimum choice for fixing the RG flow parameter $\tau$. For 
$Q\rightarrow 0$ in thermal equilibrium Eq.\ (\ref{C39}) reduces to the well
known flow parameter conditions of the previous theories \cite{D1} above
and below $T_\lambda$. The integral (\ref{C28}) can be evaluated 
asymptotically for large positive and negative $\zeta$. For $\zeta \gg +1$ 
we find 
\begin{equation}
{\cal F}_\alpha(\zeta) \approx  \Gamma(\alpha)\, \zeta^{-\alpha} 
\label{C40}
\end{equation}
which implies $\Phi_\alpha(X) \approx 1 $. Consequently, from 
(\ref{C30})-(\ref{C33}) we obtain the amplitudes 
\begin{eqnarray}
A &\approx& \epsilon^{-1} [ \rho_1^{-\epsilon/2} - 1 ] \ ,  \label{C41}  \\ 
A_1 &\approx& \epsilon^{-1} [ (1-\epsilon/2)\, \rho_1^{-\epsilon/2} - 1 ]
\ .  \label{C42}
\end{eqnarray}
The flow parameter equation (\ref{C39}) reduces to $\rho_1(1-16u[\tau]A)=1$ 
which implies
\begin{equation}
r(l)/(\mu l)^2 = \rho = \rho_1 = 1 \ .
\label{C43}
\end{equation}
Eq.\ (\ref{C22}) implies $T>T_\lambda$ and $\tau = (T-T_\lambda)/T_\lambda$ so
that the RG flow parameter $\tau$ is identified by the reduced temperature. 
Indeed, Eq.\ (\ref{C43}) is the flow-parameter equation of the previous 
theories \cite{D1} in thermal equilibrium for $T>T_\lambda$. On the other 
hand for $\zeta\ll -1$ we find 
\begin{equation}
{\cal F}_\alpha(\zeta) \approx (\pi/3)^{1/2} (-\zeta/3)^{\alpha/2-3/4} 
\exp \{2(-\zeta/3)^{3/2} \}
\label{C44}
\end{equation}
which is exponentially large. Consequently, the amplitudes $A$ and $A_1$ 
are exponentially large so that in (\ref{C34}) and in the flow parameter
condition (\ref{C39}) only the last terms are relevant. Eliminating $A$ 
we obtain 
\begin{equation}
-2 r(l)/(\mu l)^2 = -2 \rho = 1 
\label{C45}
\end{equation}
which is the flow-parameter equation of the previous theories for 
$T<T_\lambda$. Eq.\ (\ref{C22}) implies $\tau = -2 (T-T_\lambda)/T_\lambda$
and $T<T_\lambda$. We conclude that our present theory for nonzero $Q$ 
reduces to the previous theories \cite{D1} for $\zeta\gg +1$ in the normal
fluid region well above $T_\lambda$ and for $\zeta\ll-1$ in the superfluid
region well below $T_\lambda$.

We have derived seven equations given by (\ref{C29}), (\ref{C31}), 
(\ref{C33}), (\ref{C35})-(\ref{C37}), and (\ref{C39}) which we have published 
already for $d=3$ and $\epsilon=1$ in a rapid communication \cite{H1}. These 
equations contain seven variables $\zeta$, $\sigma$, $\tau$, $A$, $A_1$, 
$\Delta\rho^\prime$, $\rho_1^\prime$ which can be determined uniquely by 
solving the equations supposed the temperature difference $\Delta T=T-
T_\lambda$ and the heat current $Q$ are known. The remaining effective 
parameters, which we have eliminated, can be determined afterwards. In 
practice we have solved the equations for $d=3$ dimensions and $\epsilon=1$ 
in the following way. While the heat current $Q$ is assumed to be constant we 
take $\zeta$ as a variable which we vary in the whole interval $-\infty<\zeta 
<+\infty$ to scan all temperatures. Eq.\ (\ref{C39}) is solved explicitly to 
obtain $\sigma$ as a function of $\zeta$ and $\tau$. Then we solve the 
equations numerically by adjusting the flow parameter $\tau$ and eventually 
determine the temperature difference $\Delta T$ and the temperature gradient 
$\partial_z T$ as functions of $\zeta$ by (\ref{C35}) and (\ref{C38}). 

As an input we need the dimensionless renormalized couplings $u[\tau]$,
$\gamma[\tau]$, $w[\tau]=w^\prime[\tau] + i w^{\prime\prime}[\tau]$, 
$F[\tau]$, and $f[\tau]$ as functions of $\tau$ which have been determined 
by Dohm \cite{D1}. Furthermore, we need the parameter $g_0$ which is related 
to the entropy at $T_\lambda$. For liquid helium at saturated vapor pressure
this parameter is \cite{TA} $g_0=2.164\times 10^{11}\ {\rm s}^{-1}$. To
calculate the correlation length $\xi(\tau)=\xi_0 \tau^{-\nu}$ as a function
of $\tau$ we use the exponent $\nu=0.671$ and the amplitude $\xi_0=1.45\times
10^{-8}\ {\rm cm}$ which were determined experimentally in 
Refs.\ \onlinecite{TA} and \onlinecite{LS}. There are no adjustable 
parameters.

\section{Thermal conductivity} \label{S05}
\subsection{Numerical evaluation of the thermal conductivity and comparison
with experiments}
\begin{figure}
\vspace*{6.9cm}
\includegraphics{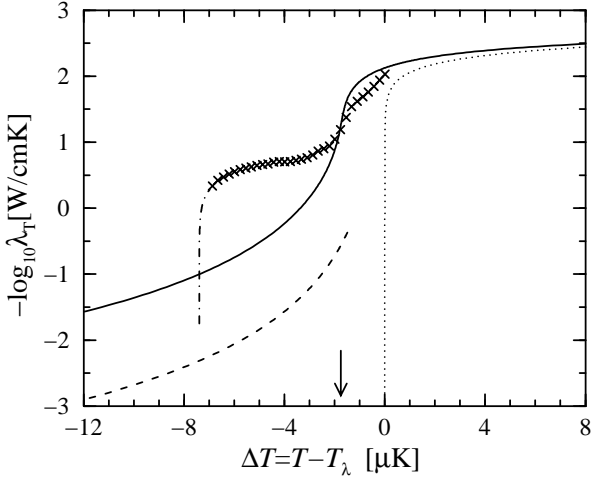}
\caption{The thermal resistivity $\rho_{\rm T}=1/\lambda_{\rm T}$ 
logarithmically as a function of $\Delta T=T-T_\lambda$ for the heat current
$Q=42.9\ \mu\mbox{W/cm}^2$. The solid line represents our theory. The data
of Liu and Ahlers \protect\cite{LA1} are shown as crosses, and the data of 
Baddar et al.\ \protect\cite{BA} (fit formula) are shown as dashed line. The 
arrow indicates $\Delta T_\lambda(Q)$. The dash-dotted line extrapolating the 
crosses indicates $\Delta T_{\rm c}(Q)$. For comparison, the thermal 
resistivity for $Q=0$ (theory) is shown as dotted line.}
\label{F03}
\end{figure}

We eliminate $\Delta\rho^\prime$ from (\ref{C36}) and (\ref{C38}) and resolve
the resulting equation with respect to $Q$. Then we obtain the heat transport
equation
\begin{equation}
Q = - \lambda_{\rm T} \, \partial_z T 
\label{D01}
\end{equation}
where
\begin{equation}
\lambda_{\rm T} = {g_0 k_{\rm B} A_d \over \tau \, \xi^{d-2} } \, 
{ \{ 1 + (f[\tau]/2) A_1 \} \over 2 \gamma[\tau] F[\tau] }  
\label{D02}
\end{equation}
is the thermal conductivity. Inserting the dimensionless parameters into
(\ref{D02}), which we calculate for given $\Delta T$ and $Q$ as described 
above by solving the seven equations, we obtain the thermal conductivity 
$\lambda_{\rm T}=\lambda_{\rm T}(\Delta T,Q)$ as a function of $\Delta T$
and $Q$. The result is obtained without adjustable parameters. We plot the 
thermal resistivity $\rho_{\rm T}=1/\lambda_{\rm T}$ logarithmically as a 
function of $\Delta T=T-T_\lambda$ for given heat currents $Q$. In 
Fig.\ \ref{F03} our result is shown for $Q=42.9\ \mu \mbox{W/cm}^2$ as solid 
line. First of all we find that $\rho_{\rm T}$ is nonzero and 
$\lambda_{\rm T}$ is finite for all temperatures above and below $T_\lambda$. 
For $T$ well above $T_\lambda$ and well below $T_\lambda$ we approximately 
find asymptotic power laws, which we will discuss in the next subsection. 

In the intermediate region, where $T$ is close to $T_\lambda$, the curve 
interpolates the two approximate power laws and shows a point of maximum
slope. We may interpret this point as the superfluid transition and define 
the related shift of the critical temperature $\Delta T_\lambda(Q)$, which 
in Fig.\ \ref{F03} is indicated by the arrow. In the previous theory 
\cite{HD2} the formula 
\begin{equation}
\Delta T_\lambda(Q) = - M\, T_\lambda \Bigl( {Q\, \xi_0^{d-1} 
\over g_0 k_{\rm B} T_\lambda} \Bigr)^x 
\label{D03}
\end{equation}
was derived for the shift of the critical temperature with the exponent 
$x=[(d-1)\nu]^{-1}=0.745$ and the constant $M=2.90$ for $d=3$ dimensions.
Here we use (\ref{D03}) as a fit formula for the point of maximum slope in
Fig.\ \ref{F03}. By varying the heat current $Q$ we find nearly the same 
exponent $x=0.745$ where deviations due to nonasymptotic effects of the
dynamic RG theory are very small here. Furthermore, we find the constant
$M=3.17$ which also is nearly the same. Thus, the point of maximum slope in
Fig.\ \ref{F03} may indeed be identified as the superfluid transition. 
However, in contrast to the previous theory \cite{HD2}, here $\Delta 
T_\lambda (Q)$ and the superfluid transition are not sharply defined for 
nonzero $Q$ because the curves of the physical quantities are smooth. 

For zero heat current (linear response limit) the thermal conductivity is
shown as dotted line in Fig.\ \ref{F03}, which is finite for $\Delta T>0$,
diverges at $\Delta T=0$, and is infinite for $\Delta T<0$. Clearly, the
solid line of our present theory ($\lambda_{\rm T}(\Delta T,Q)$ at nonzero
$Q$) approaches the dotted line ($\lambda_{\rm T}$ at $Q=0$) asymptotically
in the normal fluid region for large positive $\Delta T$. Furthermore, the
present theory reproduces the previous theory of Ref.\ \onlinecite{HD1} for
$\Delta T \gtrsim 0$. Some small deviations of the two approaches from each
other occur because here and in Ref.\ \onlinecite{HD1} the RG flow parameter
$\tau$ is determined by different conditions (the condition (\ref{C39}) here
differs from (4.49) in the second paper of Ref.\ \onlinecite{HD1}). However, 
these deviations are within the errors of the RG theory and hence are not 
serious. 

Liu and Ahlers \cite{LA1} considered a vertical heat flow upwards in liquid 
$^4$He. They measured the temperatures at the bottom and at the top of the
cell and determined the thermal conductivity $\lambda_{\rm T}(\Delta T,Q)$
in two different ways. First, they used a fit formula with power laws for
$\lambda_{\rm T}(\Delta T,Q)$ and adjusted the exponents and amplitudes.
Secondly, they obtained $\lambda_{\rm T}$ by the differential formula (24)
of the first paper of Ref.\ \onlinecite{HD1}. In Fig.\ \ref{F03} the data
obtained by the differential formula are shown as crosses. For lower 
temperatures the data are extrapolated by the fit formula shown as dash-dotted
line. While $\Delta T_\lambda(Q)$ is indicated by the arrow, the data show 
the second transition temperature $\Delta T_{\rm c}(Q)$ where the dash-dotted
line drops down nearly vertically. In the normal fluid region for $\Delta T 
\geq \Delta T_\lambda(Q)$ the experimental data agree with our theoretical 
prediction (solid line) within the accuracies of theory and experiment. 
However, in the superfluid region for $\Delta T \leq \Delta T_\lambda(Q)$ 
the data of Liu and Ahlers \cite{LA1} do not agree with our theory. For
temperatures $\Delta T$ in the interval $\Delta T_{\rm c}(Q) \leq \Delta T
\leq \Delta T_\lambda(Q)$, the so called dissipative region \cite{LA1},
the experiment finds a much larger thermal resistivity $\rho_{\rm T}=1/
\lambda_{\rm T}$, by about a factor of $20$, than our theory predicts. Then
suddenly, when the temperature $\Delta T$ approaches and drops below the 
second transition temperature $\Delta T_{\rm c}(Q)$, the experimentally 
observed thermal resistivity is so small, that it was not detected any more 
in the experiment \cite{LA1}. Our theory does not predict the second 
transition temperature $\Delta T_{\rm c}(Q)$, which has been found in the 
experiments \cite{DAS,LA1,MM}. While the solid line in Fig.\ \ref{F03} has a 
inflection point with maximum slope at $\Delta T_\lambda(Q)$, nothing unusual 
is found at $\Delta T_{\rm c}(Q)$. 

The experiments of Refs.\ \onlinecite{LA1} and \onlinecite{MM} were performed 
by measuring the temperatures at the bottom plate and at the top plate, where 
the heat current flows into and out off the helium. For this reason, these
experiments may be influenced considerably by surface effects. In superfluid 
$^4$He the dissipation of the heat current is caused by creation of vortices,
where the thermal resistivity $\rho_{\rm T}=1/\lambda_{\rm T}$ is 
proportional to the density of vortices in the helium. Near the bottom and
top surfaces additional vortices may be created which enhance the vortex
density there. This effect may possibly explain the strong enhancement of
the experimentally observed \cite{LA1} thermal resistivity for temperatures
$\Delta T$ in the interval $\Delta T_{\rm c}(Q) \leq \Delta T \leq \Delta
T_\lambda(Q)$ (see crosses in Fig.\ \ref{F03}). 

In a recent experiment Baddar et al.\ \cite{BA} measured the temperature
gradient $\partial_z T$ in superfluid helium for several heat currents $Q$. 
To exclude surface effects at the bottom and top plates the temperatures 
were measured by sidewall thermometers only. The thermal conductivity 
$\lambda_{\rm T}(\Delta T,Q)$ was then obtained from the heat transport 
equation (\ref{D01}). A power-law fit formula for $\lambda_{\rm T}
(\Delta T,Q)$ was found which is valid for a wide range of heat currents $Q$ 
and temperatures $\Delta T$ sufficiently below $\Delta T_\lambda(Q)$. In
Fig.\ \ref{F03} the thermal resistivity $\rho_{\rm T}=1/\lambda_{\rm T}$ for
$Q=42.9\ \mu\mbox{W/cm}^2$ represented by this fit formula is shown as
dashed line. Clearly, the temperature dependence of the experimental data 
agrees qualitatively with the theoretical prediction (solid line). However, 
the absolute values of the measured thermal resistivity are about a factor 
of $20$ smaller than the theoretically predicted values. While the new 
experiment of Baddar et al.\ \cite{BA} is believed to be a better and 
more direct measurement of the thermal conductivity or resistivity, there 
remains a disagreement between experiment and theory. 

We may possibly explain the discrepancy in the following way. In superfluid
$^4$He a homogeneous heat current $Q$ represents a metastable state 
\cite{HD2}. For the creation of vortices energy barriers must be overcome. 
This fact keeps the rate of vortex creation low so that the vortex density
and hence the thermal resistivity are small. On the other hand our theory is
based on the approximation where the complex order parameter $\psi$ is 
replaced by a vector $\Psi=(\psi_1,\ldots,\psi_n)$ of $n$ complex components
in the limit $n\rightarrow\infty$. In this limit the heat current $Q$ is
always unstable so that the rate of vortex creation is higher. Consequently,
in our theory the vortex density and hence the thermal resistivity are 
expected to be larger. However, the large discrepancies in Fig.\ \ref{F03}
indicate that the vortex density is a very sensitive quantity which may be
influenced strongly by the kind of the approximation in theory and by 
certain conditions in the experiment. 

\subsection{Asymptotic formulas for the thermal conductivity}
For temperatures $T$ well above $T_\lambda$ Eqs.\ (\ref{C41})-(\ref{C43}) 
imply $A=0$ and $A_1=-1/2$ so that $\lambda_{\rm T}$ depends only on 
$\tau=\Delta T/T_\lambda$ but not on the heat current. Thus, well above 
$T_\lambda$ the heat transport described by (\ref{D01}) is linear. From 
(\ref{D02}) we recover the well known result of Ref.\ \onlinecite{D1} for 
$\lambda_{\rm T}$ at infinitesimal $Q$ in one-loop order. Asymptotically in 
leading order we find
\begin{equation}
\lambda_{\rm T} \sim \tau^{-1} \xi^{-(d-2)} \sim \Delta T^{-1+(d-2)\nu}
\label{D04}
\end{equation}
which diverges in the limit $\Delta T\rightarrow 0$. For $d=3$ including
nonasymptotic effects of the dynamic RG theory one obtains $\lambda_{\rm T} 
\sim \Delta T^{-0.44}$. Thus, near $T_\lambda$ the thermal conductivity is 
strongly enhanced by critical fluctuations. 

However, at fixed nonzero $Q$ for temperatures $T$ close to $T_\lambda$ the 
flow parameter $\tau$ will reach a minimum value and the correlation length 
$\xi$ will reach a maximum value, so that $\lambda_{\rm T}$ remains finite 
even at $T_\lambda$. This fact was found previously in Ref.\ \onlinecite{HD1} 
and is seen clearly in Fig.\ \ref{F03}. The thermal conductivity 
$\lambda_{\rm T}=\lambda_{\rm T}(\Delta T,Q)$ becomes $Q$ dependent so that 
the heat transport is nonlinear. The crossover from linear to nonlinear heat 
transport happens for temperatures $\Delta T$ below 
\cite{HD1} 
\begin{equation}
\Delta T_{\rm nl}(Q) = M_{\rm nl}\, T_\lambda \Bigl( {Q\, \xi_0^{d-1} 
\over g_0 k_{\rm B} T_\lambda} \Bigr)^x 
\label{D05}
\end{equation}
with the exponent $x=[(d-1)\nu]^{-1}$, where $M_{\rm nl}$ is a constant of 
order unity. For $d=3$ the values $x=0.745$ and $M_{\rm nl}\approx 2.8$ were
found \cite{HD1}.

While the previous theory \cite{HD1} is valid only for $\Delta T\gtrsim 0$, 
the present theory works also for lower temperatures in the superfluid region.
Well below $T_\lambda$ the thermal conductivity (\ref{D02}) can be 
evaluated asymptotically. For $\zeta\lesssim -5$ the function 
${\cal F}_\alpha(\zeta)$ approximated by (\ref{C44}) is exponentially large. 
Consequently, the amplitudes $A$ and $A_1$ are exponentially large. From
(\ref{C31}) and (\ref{C33}) we obtain the ratio
\begin{equation}
{A_1 \over A} \approx - { \zeta \, {\cal F}_{\epsilon/2}(\zeta) \over
{\cal F}_{-1+\epsilon/2}(\zeta) }  \approx  3^{-1/2} (-\zeta)^{3/2}  \ .
\label{D06}
\end{equation}
Eqs.\ (\ref{C36}) and (\ref{C37}) reduce to
\begin{eqnarray}
{F[\tau] \, \Delta\rho^\prime \over 4 \gamma[\tau] w^\prime[\tau] }
&\approx& - {1\over A_d} \Bigl( {Q \xi^{d-1} \over g_0 k_{\rm B} T_\lambda }
\Bigr) \ {1\over A_1} \ ,  \label{D07}  \\
\rho_1^\prime &\approx& - {1\over \tau} \, {\xi\over T_\lambda} \, 
{\partial T_\lambda \over \partial z} \, {1\over 8 u[\tau]} \ {1\over A_1} \ . 
\label{D08}
\end{eqnarray}
We assume that the heat current $Q$ is sufficiently large so that gravity 
effects may be neglected and (\ref{D08}) is much smaller than (\ref{D07}).
Then, from (\ref{C29}) we obtain 
\begin{equation}
\sigma \approx {1\over 12 A_d^2} \, \Bigl( {Q \xi^{d-1} \over g_0 k_{\rm B} 
T_\lambda } \Bigr)^2  \, {1\over A_1^2} \ . 
\label{D09}
\end{equation}
On the other hand, the flow parameter equation (\ref{C39}) reduces to
\begin{equation}
\sigma^{1/3} 16 u[\tau] A \, (-\zeta) = 1 \ . 
\label{D10}
\end{equation}
Now, Eqs.\ (\ref{D06}), (\ref{D09}), and (\ref{D10}) are three equations for
$A$, $A_1$, and $\sigma$. Eliminating $\sigma$ we obtain the amplitudes 
\begin{eqnarray}
A &\approx& {1\over 2} \, {A_d^2 \over (8u[\tau])^3} 
\Bigl( {g_0 k_{\rm B} T_\lambda \over Q \xi^{d-1} } \Bigr)^2 \ ,  
\label{D11} \\
A_1 &\approx& {(-\zeta)^{3/2}\over 2\sqrt{3}} \, {A_d^2 \over (8u[\tau])^3} 
\Bigl( { g_0 k_{\rm B} T_\lambda \over Q \xi^{d-1} } \Bigr)^2 \ . 
\label{D12}
\end{eqnarray}
Eventually, from (\ref{D02}) we obtain the thermal conductivity 
\begin{equation}
\lambda_{\rm T} \approx {g_0 k_{\rm B} \over \tau\, \xi^{d-2}} \,
{F[\tau] \over 4\gamma[\tau] w^\prime[\tau]} \, 
{1\over \sqrt{12}} \, \Bigl( {(-\zeta)^{1/2} A_d \over 8u[\tau]} \Bigr)^3
\Bigl( { g_0 k_{\rm B} T_\lambda \over Q \xi^{d-1} } \Bigr)^2 , 
\label{D13}
\end{equation}
where $\tau= -2\Delta T/T_\lambda$. The variable $\zeta$ depends only weakly
on $\tau$, i.e. logarithmically. Thus, asymptotically in leading order we find 
\begin{equation}
\lambda_{\rm T} \sim \tau^{-1} \xi^{4-3d} Q^{-2} \sim 
(-\Delta T)^{(3d-4)\nu-1} Q^{-2} \ . 
\label{D14}
\end{equation}
For $d=3$ including nonasymptotic effects we obtain $\lambda_{\rm T} \sim
(-\Delta T)^{2.4} Q^{-2}$. 

In their experiment Baddar et al. \cite{BA} have found that for a wide 
range of heat currents $Q$ the thermal conductivity can be expressed in
terms of the power-law formula 
\begin{eqnarray}
\lambda_{\rm T,exp} &=& \lambda_0 \bigl[ (-\Delta T/T_\lambda) 
(Q/Q_0)^{-0.904} \bigr]^{2.8}  \label{D15} \\
&\sim & (-\Delta T)^{2.8} Q^{-2.53}  \label{D16}
\end{eqnarray}
where $\lambda_0=1\ \mbox{W/cmK}$ and $Q_0=393\ \mbox{W/cm}^2$. Clearly,
this power-law formula has the same structure as the asymptotic formula 
(\ref{D14}) of our theoretical prediction. However, the two formulas do not 
agree quantitatively with each other. The exponents of the power laws differ
by about $20\%$. Furthermore, the amplitudes differ by a factor of $20$ 
which means that $\lambda_{\rm T,exp}$ is about $20$ times larger than the 
theoretical $\lambda_{\rm T}$ (see dashed and solid line in Fig.\ \ref{F03}). 
The discrepancies may possibly be due to the approximation we have used in
our theory. Further theoretical and experimental work is necessary to clarify
the origin of the discrepancies. 

\subsection{Influence of gravity}
On earth gravity implies a spatially dependent superfluid transition 
temperature $T_\lambda(z)$ with a nonzero gradient $\partial_z T_\lambda=
\pm 1.273\ \mu\mbox{K/cm}$. Supposed the heat current $Q$ is flowing in the
$z$ direction the gradient $\partial_z T_\lambda$ is positive for heat 
current upwards and negative for heat current downwards. In our theory the
gradient $\partial_z T_\lambda$ is incorporated in (\ref{C24}) or equivalently
in (\ref{C37}). We find that on earth the effects of heat current $Q$ and of
gravity have equal magnitude for $Q\approx 65\ \mbox{nW/cm}^2$. For larger
$Q$ the heat current is dominating while for smaller $Q$ gravity is 
dominating. We find that for $Q\gtrsim 1\ \mu\mbox{W/cm}^2$ the effects of
gravity are very small so that in this case gravity can be neglected. The
experiments by Liu and Ahlers \cite{LA1}, Murphy and Meyer \cite{MM}, and the
new experiment by Baddar et al. \cite{BA} are performed at heat currents $Q$
which satisfy this condition. 

We have performed our calculations for a positive and a negative gradient 
$\partial_z T_\lambda$ representing gravity on earth and also for a zero 
gradient $\partial_z T_\lambda$ which corresponds to a microgravity 
environment in space. As a result, for heat currents $Q\gtrsim 100
\ \mbox{nW/cm}^2$ the thermal conductivity $\lambda_{\rm T}(\Delta T,Q)$ is 
nearly independent of gravity. On the other hand, for heat currents 
$Q\lesssim 80\ \mbox{nW/cm}^2$ our theory fails if gravity is present, because 
$\sigma$ defined in (\ref{C29}) changes sign so that no solution of the seven 
equations for the dimensionless parameters in Sec. \ref{S04} can be found. 

Day et al. \cite{DMM} investigated the superfluid-normal-fluid interface
in $^4$He and measured the thermal conductivity $\lambda_{\rm T}$ for very
small heat currents $Q$ in the interval $20\ \mbox{nW/cm}^2 < Q < 6\ \mu
\mbox{W/cm}^2$. The heat current flows upwards so that the gradient 
$\partial_z T_\lambda$ is positive. Clearly, this experiment explores the 
crossover from the gravity dominated region to the heat current dominated 
region. For heat currents $Q\gtrsim 100\ \mbox{nW/cm}^2$ the experimental
data agree quite well with our theoretical prediction for $\lambda_{\rm T}$.
This has been demonstrated in our previous rapid communication \cite{H1} for
temperatures $\Delta T\gtrsim \Delta T_\lambda(Q)$ (see Fig.\ 2 therein). 
In the superfluid region for $\Delta T\lesssim \Delta T_\lambda(Q)$ the 
thermal resistivity is very small so that here experimental data are not
available with sufficient accuracy on a logarithmic scale. 

For very low heat currents $Q\leq 40\ \mbox{nW/cm}^2$ the data of Day et 
al. \cite{DMM} indicate that the thermal resistivity $\rho_{\rm T} = 1/
\lambda_{\rm T}(\Delta T,Q)$ is a smooth function of $\Delta T$ in the limit
$Q\rightarrow 0$ if gravity is present. This fact means that gravity prevents
the system from reaching the critical point of the superfluid transition, so 
that all physical quantities are smooth and nonsingular near $T_\lambda$. 
Unfortunately, our theory fails for these low heat currents in the gravity
dominated region. However, an alternative approach is possible which is
equivalent to Onuki's theory \cite{O1}. In mean-field approximation the
model-{\it F\,} equations (\ref{A01}) and (\ref{A02}) can be solved numerically
as partial differential equations to obtain the order-parameter profile and
the temperature profile of the superfluid-normal-fluid interface. To include
the critical fluctuations the model-{\it F\,} equations are renormalized and
the RG theory is applied. The approach is a renormalized mean-field theory
which works for all heat currents $Q$ and gravity, even for very small $Q$ 
in the gravity dominated region. In this way we obtain a thermal resistivity
$\rho_{\rm T}=1/\lambda_{\rm T}(\Delta T,Q)$ which agrees qualitatively with
the experimental observation of Day et al. \cite {DMM}: it is a smooth 
function of $\Delta T$ and nearly independent of $Q$ for $Q\lesssim 
10\ \mbox{nW/cm}^2$ including the limit $Q\rightarrow 0$. In agreement with
the experiment \cite{DMM} we find that $\delta T_g \approx 15\ \mbox{nK}$ is 
the relevant temperature scale on which the critical singularity is 
smoothened by gravity. This temperature scale is related to the thickness of 
the superfluid-normal-fluid interface $\xi_g \approx 100\ \mu\mbox{m}$ by 
$\delta T_g/\xi_g \approx \vert \partial_z T_\lambda \vert = 1.273\ \mu
\mbox{K/cm}$.

To measure the critical singularity of the thermal resistivity or 
conductivity for heat currents smaller than $100\ \mbox{nW/cm}^2$ and 
temperatures closer than $15\ \mbox{nK}$ to the superfluid transition, the 
experiment must be performed under microgravity conditions in space. Of 
course, for zero gravity where $\partial_z T_\lambda=0$ our theory presented 
in this paper works for all heat currents and never fails at any low $Q$.

\section{The temperature profile} \label{S06} 
\subsection{Numerical results}
\begin{figure}[t]
\vspace*{6.9cm}
\includegraphics{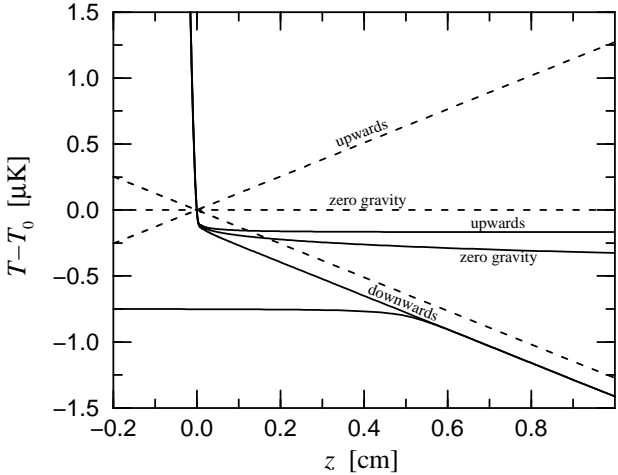}
\caption{The temperature profile of the superfluid-nor\-mal-fluid interface 
for the heat current $Q=1\ \mu \mbox{W/cm}^2$ for the three 
gravity conditions: vertical heat flow {\it upwards} and {\it downwards} on 
earth and heat flow in {\it zero gravity} in space. While the solid lines
correspond to the temperature profiles $T(z)$, the dashed lines represent the
spatially dependent critical temperatures $T_\lambda(z)$, which reflect the 
gravity conditions.}
\label{F05}
\end{figure}

Once the thermal conductivity $\lambda_{\rm T}=\lambda_{\rm T}(\Delta T,Q)$
has been determined, the temperature profile $T(z)$ is calculated by 
solving the heat transport equation (\ref{D01}) as a differential equation 
which can be written in the form
\begin{equation}
{\partial T\over \partial z}  = - {Q \over \lambda_{\rm T}(\Delta T,Q)} 
\label{E01}
\end{equation}
where $\Delta T=T(z)-T_\lambda(z)$. In Fig.\ \ref{F05} the resulting 
temperature profile of the superfluid-normal-fluid interface is shown for 
$Q=1\ \mu\mbox{W/cm}^2$. The several curves correspond to all three gravity
conditions as indicated in the figure: vertical heat flow {\it upwards} and 
vertical heat flow {\it downwards} on earth and heat flow in {\it zero 
gravity} in space. The temperature profiles $T(z)$ are shown as solid 
lines. The spatially dependent critical temperatures $T_\lambda(z)$, which 
represent the gravity conditions, are shown as dashed lines. While in 
Sec.\ \ref{S05} we have found that $\lambda_{\rm T}(\Delta T,Q)$ does not 
depend on gravity for $Q\gtrsim 0.1\ \mu\mbox{W/cm}^2$, the temperature 
profile $T(z)$ is considerably influenced by gravity which is clearly seen 
in Fig.\ \ref{F05} because there are several solid lines. 

In Fig.\ \ref{F05} we have chosen the coordinates so that the 
superfluid-normal-fluid interface, where $\Delta T=T(z)-T_\lambda(z)=0$, is
located at $z=0$ and $T(z)=T_0$. For this reason all curves intersect with
each other in the point $(0,0)$. For $z<0$ the helium is normal fluid because
$\Delta T=T(z)-T_\lambda(z)>0$ so that the solid line is above the respective 
dashed line. On the other hand for $z>0$ the helium is superfluid because 
here $\Delta T=T(z)-T_\lambda(z)<0$ so that the respective solid line is
below the respective dashed line. (Since the many curves in Fig.\ \ref{F05}
may be confusing, one should have in mind that only those solid and dashed
lines should be compared with each other which belong to the same gravity
condition.) Furthermore, we have chosen a strongly enlarged temperature scale 
(in $\mu\mbox{K}$) which resolves the very small temperature gradients in the
superfluid region $z>0$. For this reason in the normal-fluid region $z<0$ the
solid line in Fig.\ \ref{F05} (temperature profile $T(z)$\,) has a very steep
gradient and goes up nearly vertically for decreasing $z$. The shifted 
critical temperature $\Delta T_\lambda(Q)$, which here is $-0.10\ \mu
\mbox{K}$, is located approximately at this point where the solid line
has a ``round corner'' and the gradient changes from large negative to small
negative. 

In the normal-fluid region $z\lesssim 0$ where $\Delta T\gtrsim \Delta
T_\lambda(Q)$ the three solid lines fall all together into one line which 
indicates that here gravity is negligible. On the other hand in the superfluid
region $z\gtrsim 0$ where $\Delta T\lesssim \Delta T_\lambda(Q)$ all three
solid lines differ from each other which means that here gravity is important.
A general criterion can be found to distinguish the regions where gravity is
important and where it is negligible. In the differential equation (\ref{E01})
gravity is included implicitly via $\Delta T=T(z)-T_\lambda(z)$ by the 
gradient of the critical temperature $T_\lambda(z)$. Gravity will be important
or not if the spatial dependence of $T_\lambda(z)$ or of $T(z)$ is 
dominating. We may define the $Q$-dependent temperature difference $\Delta
T_1(Q)$ by the condition $\vert\bbox{\nabla}T\vert= \vert\bbox{\nabla}
T_\lambda\vert$, which in terms of (\ref{E01}) can be written as
\begin{equation}
{Q \over \lambda_{\rm T}(\Delta T_1(Q),Q)} = \Bigl\vert {\partial T_\lambda 
\over \partial z} \Bigr\vert = 1.273\ \mu\mbox{K/cm}  \ . 
\label{E02}
\end{equation}
For $\Delta T>\Delta T_1(Q)$ it is $\vert\bbox{\nabla}T\vert> \vert
\bbox{\nabla} T_\lambda\vert$ so that the heat current is dominating. On the
other hand for $\Delta T<\Delta T_1(Q)$ it is $\vert\bbox{\nabla}T\vert< 
\vert\bbox{\nabla} T_\lambda\vert$ so that gravity is dominating. 
Consequently, for temperatures $\Delta T$ sufficiently well above $\Delta 
T_1(Q)$ it is $\vert\bbox{\nabla}T\vert\ll \vert \bbox{\nabla} T_\lambda\vert$ 
so that gravity can be neglected while otherwise gravity effects are
significant. Thus, $\Delta T_1(Q)$ may be viewed as the temperature which
separates the heat current dominated region from the gravity dominated region.

\begin{figure}[t]
\vspace*{6.9cm}
\includegraphics{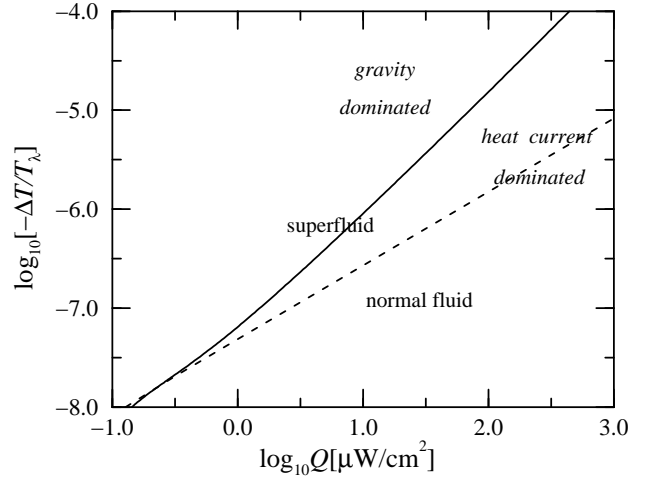}
\caption{The temperature shifts $\Delta T_1(Q)$ (solid line) and 
$\Delta T_\lambda(Q)$ (dashed line), obtained from (\ref{E02}) and 
(\ref{D03}), respectively, as functions of the heat current $Q$ in a 
double logarithmic plot. $\Delta T_1(Q)$, separates the gravity dominated 
region from the heat current dominated region, while $\Delta T_\lambda(Q)$ 
separates the superfluid region from the normal-fluid region.}
\label{F06}
\end{figure}

For the heat current $Q=1\ \mu\mbox{W/cm}^2$ in Fig.\ \ref{F05} we find 
$\Delta T_1(Q)=-0.14\ \mu\mbox{K}$ which is slightly below but close to 
$\Delta T_\lambda(Q)=-0.10\ \mu\mbox{K}$. Thus, in this case the interface 
between the superfluid and the normal-fluid region nearly coincides with the
interface which separates the gravity dominated region from the heat current
dominated region. This fact is clearly seen in Fig.\ \ref{F05}. For other 
heat currents $Q$ the situation may change, because $\Delta T_1(Q)$ and
$\Delta T_\lambda(Q)$ may be farther apart from each other. In Fig.\ \ref{F06}
we plot $\Delta T_1(Q)$ and $\Delta T_\lambda(Q)$ as functions of $Q$ on a
double logarithmic scale. Our theoretical result for $\Delta T_1(Q)$ obtained
from (\ref{E02}) is shown as solid line. Furthermore, $\Delta T_\lambda(Q)$
obtained from (\ref{D03}) is shown as dashed line. For the heat currents
$Q\gtrsim 0.1\ \mu\mbox{W/cm}^2$, for which our theory is valid in gravity,
both $\Delta T_\lambda(Q)$ and $\Delta T_1(Q)$ are negative. While in 
Fig.\ \ref{F06} the solid line represents $\Delta T_1(Q)$ for vertical heat
flow downwards, $\Delta T_1(Q)$ for vertical heat flow upwards will be
slightly different. However, the difference is very small. It is smaller 
than the width of the solid line so that it can be neglected. Thus, the solid
line in Fig.\ \ref{F06} represents $\Delta T_1(Q)$ for both heat flow 
directions with sufficient accuracy.

For $Q=0.22\ \mu\mbox{W/cm}^2$, in Fig.\ \ref{F06} the solid line and the 
dashed line intersect each other so that $\Delta T_1(Q)=\Delta T_\lambda(Q)$.
For larger heat currents $\Delta T_1(Q)$ is always below $\Delta T_\lambda(Q)$
so that the separation between the gravity dominated and the heat current
dominated region is located always in the superfluid region or equivalently
the superfluid-normal-fluid interface is located in the heat current 
dominated region. It turns out that for sufficiently large heat currents,
say $Q\gtrsim 10\ \mu\mbox{W/cm}^2$, the superfluid-normal-fluid interface
is nearly free from influences of gravity, while gravity effects rise to a
significant magnitude only far away from the interface in the superfluid
region. However, since the interface thickness must be larger than the size
of the thermometers, experiments to resolve the temperature profile of the
superfluid-normal-fluid interface must be performed for very small heat
currents $Q\lesssim 0.1\ \mu\mbox{W/cm}^2$. For these small heat currents,
on earth the interface would be strongly influenced by gravity. For this
reason, an experiment to measure the temperature profile of the interface
is prepared to be performed under microgravity conditions in space \cite{DY}. 

\subsection{Asymptotic formulas}
By using the asymptotic formulas of the thermal conductivity $\lambda_{\rm T}
(\Delta T,Q)$ in Sec.\ \ref{S05}.B, the differential equation (\ref{E01})
can be solved explicitly so that asymptotic formulas for the temperature
profile $T(z)$ are found in several regions. First of all we consider the
heat flow in zero gravity where $T_\lambda(z)=T_\lambda$ is constant. In this
case the differential equation (\ref{E01}) can be integrated easily by 
separation of the variables so that
\begin{equation}
z\ =\ - \int_{T_\lambda}^T Q^{-1} \ \lambda_{\rm T}(T^\prime -T_\lambda,Q) 
\ dT^\prime \ . 
\label{E03}
\end{equation}
In the normal fluid region for $T$ sufficiently well above $T_\lambda$ the
asymptotic formula (\ref{D04}) may be inserted. Consequently, we obtain
\begin{equation}
\Delta T(z) = T(z)-T_\lambda \approx K_+ (-Qz)^{1/[(d-2)\nu]} 
\label{E04}
\end{equation}
for $z\ll 0$ with a certain constant $K_+$. On the other hand, for $T$ 
sufficiently well below $T_\lambda$ the asymptotic formula (\ref{D14}) may
be inserted which implies
\begin{equation}
\Delta T(z) = T(z)-T_\lambda \approx - K_- (Q^3z)^{1/[(3d-4)\nu]} 
\label{E05}
\end{equation}
for $z\gg 0$ with a certain constant $K_-$. Including the nonasymptotic
effects of the dynamic RG theory, for the temperature profile $T(z)$ we
obtain the asymptotic formula 
\begin{equation}
T(z)\approx \cases{ T_\lambda  + K_+ (-Qz)^{1.79} &for $z\ll 0$ \cr
T_\lambda  - K_- (Q^3z)^{0.294} &for $z\gg0$ \cr }
\label{E06}
\end{equation}
which is valid for the heat flow in zero gravity.  This formula can be 
compared with the solid line in Fig.\ \ref{F05} for zero gravity. Clearly,
the temperature profile $T(z)$ decreases monotonically with increasing $z$.
There is no lower bound for $T(z)$ in the limit $z\rightarrow\infty$. 

Asymptotic formulas for the temperature profile $T(z)$ can be found also 
for vertical heat flows in gravity. The interface, which separates the
heat current dominated region from the gravity dominated region, is located
at a coordinate $z_1$ defined by $\Delta T(z_1) = \Delta T_1(Q)$. For 
$Q>0.22\ \mu\mbox{W/cm}^2$ it is $z_1>0$ so that the interface is located in
the superfluid region. For $z\ll z_1$ gravity is negligible. Consequently,
the asymptotic formula (\ref{E06}) remains valid for $z\ll z_1$ where
$T_\lambda=T_\lambda(z=0)$ should be inserted. In the case of a vertical
heat flow {\it upwards} another asymptotic formula can be found for 
$z\gg z_1$. In Fig.\ \ref{F05} it is clearly seen that in this case the
temperature profile approaches a limiting value
\begin{equation}
\lim_{z\to +\infty} T(z) = T_\infty(Q) 
\label{E07}
\end{equation}
which is below $T_\lambda(z=0)$. For $z\gg z_1$ the temperature difference is
approximately $\Delta T(z) \approx T_\infty(Q) - T_\lambda(z)$ so that its
$z$ dependence is governed by $T_\lambda(z)$, i.e. by gravity. Then, by using
the asymptotic formula $\lambda_{\rm T}\sim (-\Delta T)^{2.4} Q^{-2}$ for
the thermal conductivity we obtain the asymptotic formula 
\begin{equation}
T(z) \approx T_\infty(Q) + K_\infty Q^3 (z-z_0)^{-1.4} \quad 
\mbox{for $z\gg z_1$} 
\label{E08}
\end{equation}
where $K_\infty>0$ and $z_0<0$ are certain constants and $T_\infty(Q)=
T_\lambda(z=0) + z_0\, \partial_z T_\lambda$. 

On the other hand, for a vertical heat flow {\it downwards} the solid
line in Fig.\ \ref{F05} approaches a straight line parallel to the dashed 
line for $z\gg z_1$ so that the gradients $\partial_z T$ and $\partial_z
T_\lambda$ are nearly equal. The distance from criticality $\Delta T(z) =
T(z)-T_\lambda(z)$ approaches a constant value given by 
\begin{equation}
\lim_{z\to +\infty} \Delta T(z) = \Delta T_1(Q)  \ .
\label{E09}
\end{equation}
Consequently, for the temperature profile we find the asymptotic formula 
\begin{equation}
T(z) \approx T_\lambda(z) + \Delta T_1(Q) \quad \mbox{for $z\gg z_1$}  \ . 
\label{E10}
\end{equation}

\subsection{Dissipative region} 
For the vertical heat flow upwards the asymptotic formulas (\ref{E06}) and
(\ref{E08}) indicate the existence of a dissipative region which may be 
related to the dissipative region observed in the experiments by Liu and 
Ahlers \cite{LA1} and by Murphy and Meyer \cite{MM}. The superfluid 
transition happens in two steps at $z=0$ and $z=z_1$ or equivalently at the
temperatures $\Delta T=\Delta T_\lambda(Q)$ and $\Delta T=\Delta T_1(Q)$. 
The region $0\lesssim z\lesssim z_1$, which corresponds to $\Delta T_1(Q)
\lesssim \Delta T \lesssim \Delta T_\lambda(Q)$ (in Fig.\ \ref{F06} the
region between the solid and the dashed line), may be identified as the 
dissipative region, because here the temperature profile $T(z)$ has a finite
gradient. On the other hand, the region $z\gtrsim z_1$ which corresponds
to $\Delta T \lesssim \Delta T_1(Q)$ (in Fig.\ \ref{F06} the region above the
solid line) is the really superfluid region, because here the asymptotic 
formula (\ref{E08}) implies a nearly flat temperature profile with a nearly
zero gradient. Thus, $\Delta T_1(Q)$ may be interpreted as the transition
temperature between the dissipative region and the really superfluid region,
which should be related to the temperature shift
\begin{equation}
\Delta T_{\rm c}(Q) = T_\infty(Q) - T_\lambda(z=0) 
\label{E11}
\end{equation}
measured in the experiments \cite{DAS,LA1,MM}. We find that $\Delta T_{\rm c}
(Q)$ obtained from our theory is close to $\Delta T_1(Q)$ and located slightly
above the solid line in Fig.\ \ref{F06}. From the slopes of the lines in the 
double-logarithmic plot we obtain effective power laws $\Delta T_1(Q) \sim
-Q^x$ and $\Delta T_{\rm c}(Q) \sim -Q^x$ with nearly the same exponent $x$,
which varies between $x\approx 0.9$ for $0.1\ \mu\mbox{W/cm}^2 \lesssim Q 
\lesssim 1\ \mu\mbox{W/cm}^2$ and $x\approx 1.25$ for $Q\gtrsim 10\ \mu
\mbox{W/cm}^2$.  

While our theoretical predictions for the dissipati\-ve region agree 
qualitatively with the experimental ob\-servations \cite{DAS,LA1,MM}, there
are three major quantitative disagreements. First of all, the experimentally 
observed dissipation is much larger than the theoretically predicted, because
for $\Delta T_{\rm c}(Q) < \Delta T < \Delta T_\lambda(Q)$ the measured 
thermal resistivity $\rho_{\rm T}=1/\lambda_{\rm T}$ of Ref.\ \onlinecite{LA1}
(crosses in Fig.\ \ref{F03}) is much larger than the theoretically predicted
(solid line in Fig.\ \ref{F03}). Secondly, the experimental $\Delta T_{\rm c}
(Q)\sim -Q^x$ does not agree with the theoretically predicted because the 
exponent $x_{\rm exp}=0.81$ of Duncan et al. \cite{DAS} is considerably 
smaller than the theoretical exponent $x\gtrsim 0.9$. Furthermore, our 
theory predicts a much larger spatial extent $\Delta z=z_1$ of the dissipative
region than it is in the experiments \cite{DAS,LA1}. We find $\Delta z \gtrsim
0.25\ \mbox{cm}$ for $Q\gtrsim 10\ \mu\mbox{W/cm}^2$ which is about the sample 
size, while the experiments find a spatial extent $\Delta z$ much smaller than 
the sample size. 

The disagreements may possibly be due to the experiments of 
Refs.\ \onlinecite{DAS,LA1,MM}, because in the new experiment by Baddar et
al. \cite{BA} the dissipative region has not been observed in this form. We
may use the experimental fit formula \cite{BA} (\ref{D15}) for the thermal
conductivity to calculate the related $\Delta T_{1,{\rm exp}}(Q)$ and 
$\Delta T_{\rm c,exp}(Q)$ by (\ref{E02}) and (\ref{E11}), respectively. As
results we obtain power laws $\sim -Q^x$ with the exponent $x_{\rm exp}=1.26$.
This exponent agrees quite well with our theoretical value $x=1.25$ for
large heat currents. If we plot $\Delta T_{\rm c,exp}(Q)$ in Fig.\ \ref{F06},
the respective line would be parallel to the solid line for $Q\gtrsim 10
\ \mu\mbox{W/cm}^2$ but located somewhat below the solid line. Thus, there
remains a quantitative disagreement which is related to the quantitative
disagreement of $\lambda_{\rm T}$. However, $\Delta T_{\rm c,exp}(Q)$ does
also not agree with the $\Delta T_{\rm c}(Q)$ of the previous experiments
\cite{DAS,LA1}. Thus, for the clarification of the disagreements also further
experimental work is necessary. 

Since the thermal conductivity $\lambda_{\rm T}(\Delta T,Q)$ (solid line in
Fig.\ \ref{F03}) is smooth and does not show any unusual behavior at 
$\Delta T_1(Q)$ and $\Delta T_{\rm c}(Q)$, these temperatures are not 
properties of the helium. Rather, $\Delta T_1(Q)$ and $\Delta T_{\rm c}(Q)$ 
are implied by gravity and occur when integrating the heat transport equation
(\ref{D01}) or (\ref{E01}) to calculate the temperature profile. Thus, we
predict that in the experiment in space \cite{DY}, where gravity is zero, 
the $\Delta T_{\rm c}(Q)$ of the experiments of Refs.\ \onlinecite{DAS},
\onlinecite{LA1}, and \onlinecite{MM} should not be observable and not be
existent, while the $\Delta T_\lambda(Q)$ is expected to be found. 

\subsection{Self organized critical state}
In gravity for vertical heat flows downwards the gradients $\partial_z T$
and $\partial_z T_\lambda$ are both negative. A situation may arise where
both gradients are equal, $\partial_z T = \partial_z T_\lambda$, so that 
$\Delta T(z)= \Delta T_1(Q)$ is constant over a larger region in space. This
state of the helium represents a self organized critical (SOC) state which 
was considered theoretically by Onuki \cite{O2} and proposed for an experiment
by Ahlers and Liu \cite{AL}. Recently, Moeur et al.\ \cite{MD} have realized
the SOC state and measured the distance from criticality $\Delta T(z)=
\Delta T_1(Q)$ as a function of the heat current $Q$ for $40\ \mbox{nW/cm}^2
<Q<6\ \mu\mbox{W/cm}^2$. The experimental result agrees quite well with
our theoretical prediction for $\Delta T_1(Q)$, while our theory does not 
include any unknown adjustable parameters. This has been demonstrated in our
previous rapid communication \cite{H1} for heat currents below 
$1.5\ \mu\mbox{W/cm}^2$ (see Fig.\ 3 therein). However, for $Q\gtrsim 
1.5\ \mu\mbox{W/cm}^2$ deviations occur: for a given distance from criticality 
$\Delta T=\Delta T_1(Q)$ the related heat current $Q$ is larger in the 
experiment than in our theory. This fact means that for $Q\gtrsim 
1.5\ \mu\mbox{W/cm}^2$ the dissipation observed in the experiment \cite{MD} 
is smaller than the dissipation predicted by our theory. 

The SOC state is stable in the following sense: the distance from criticality
$\Delta T(z)$ always converges to the constant value $\Delta T_1(Q)$ for
large $z$ according to (\ref{E09}). This fact is clearly seen in 
Fig.\ \ref{F05}. However, there are two possibilities: $\Delta T(z)$ may
converge to $\Delta T_1(Q)$ either from {\it above} or from {\it below}. In
the first case an interface between normal-fluid helium and the SOC state is
found, while in the second case an interface between superfluid helium and the 
SOC state is found. In Fig.\ \ref{F05} the temperature profiles $T(z)$ are
shown for both cases, where the superfluid-SOC interface is represented by the
lowest solid line. Both kinds of interfaces were realized in the experiment
by Moeur et al.\ \cite{MD}. 

Since $\Delta T(z)=\Delta T_1(Q)$ is constant, the SOC state is homogeneous 
in space so that it is an ideal system for theoretical and experimental 
investigations. In the Appendix we evaluate the Green's function $G({\bf r},
{\bf r}^\prime)$ and the related quantities $n_{\rm s}$ and $J_{\rm s}$ by
assuming $r_1$ and $\Delta r$ to be linear functions of the space coordinate 
${\bf r}$ given by (\ref{Z12}) and (\ref{Z13}). While in general this 
assumption implies an approximation, for the SOC state the assumption is 
{\it exactly} satisfied, because $r_1$ and $\Delta r$ are directly related to
the temperature profiles $\Delta T=T(z)-T_\lambda(z)$ and $T_\lambda(z)$ 
which are constant and linear in $z$, respectively.

\section{Correlation lengths} \label{S07}
\begin{figure}[t]
\vspace*{6.9cm}
\includegraphics{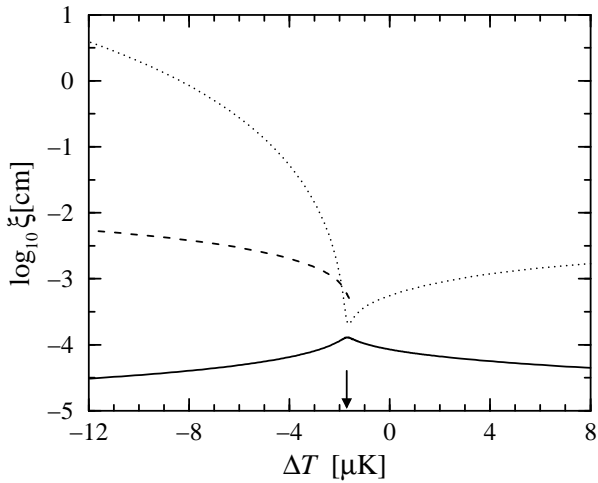}
\caption{The correlation lengths as functions of $\Delta T$ for the heat
current $Q=42.9\ \mu\mbox{W/cm}^2$. The correlation length $\xi$ is shown
as solid line. The dephasing length $\xi_1$ obtained from (\ref{G07}) is 
shown as dashed line. The dotted line represents the characteristic length 
scale $\xi_2$ of the temperature variations defined by (\ref{G11}). The 
arrow indicates $\Delta T_\lambda(Q)$.}
\label{F08}
\end{figure}

The RG theory includes a characteristic length defined by $\xi=
(\mu l)^{-1}$, which is called the correlation length. Near criticality
our theory yields the asymptotic result $\xi=\xi_0 \tau^{-\nu}$ where 
\cite{TA,LS} $\nu=0.671$ and $\xi_0=1.45\times 10^{-8}\ \mbox{cm}$. The
correlation length depends on $\Delta T$ and $Q$ indirectly via the RG
flow parameter $\tau=\tau(\Delta T,Q)$ determined in Sec.\ \ref{S04}. In
Fig.\ \ref{F08} the correlation length $\xi$ is shown logarithmically as a
function of $\Delta T$ for $Q=42.9\ \mu\mbox{W/cm}^2$, i.e.\ the same heat 
current as in Fig.\ \ref{F03}. Clearly, $\xi$ increases when $\Delta T$ 
approaches the superfluid transition. The nonzero heat current implies that 
$\xi$ is a smooth function of $\Delta T$ which has a maximum located at 
$\Delta T_\lambda(Q)$. In Fig.\ \ref{F08} the maximum is clearly shown by 
the solid line, where its position at $\Delta T_\lambda(Q)$ is indicated by 
the arrow. 

Correlation lengths can be observed in the equal-time Green's function 
$G({\bf r},{\bf r}^\prime)$ because they represent the characteristic length 
scales for the decay of the Green's function, when the separation of the two 
space points $\vert{\bf r}-{\bf r}^\prime\vert$ increases. In the normal
fluid region for $\Delta T>\Delta T_\lambda(Q)$ we find
\begin{equation}
G({\bf r},{\bf r}^\prime) \rightarrow 0 \quad \mbox{for $\vert{\bf r}-
{\bf r}^\prime\vert \gg \xi$}
\label{G01}
\end{equation}
which is valid for zero and nonzero heat currents $Q$. In the superfluid 
region for $\Delta T< \Delta T_\lambda(Q)$ the decay happens in two steps 
according to 
\begin{equation}
G({\bf r},{\bf r}^\prime) \rightarrow \cases{ \eta^2\, e^{i{\bf k}({\bf r}-
{\bf r}^\prime)} &for $\xi\ll \vert{\bf r}-{\bf r}^\prime\vert \ll \xi_1$ \cr
0 &for $\vert{\bf r}-{\bf r}^\prime\vert \gg \xi_1$ \cr}
\label{G02}
\end{equation}
where $\xi_1$ is a second correlation length which will be defined below. In 
this case the decay is strongly influenced by the heat current via ${\bf k}$
and $\xi_1$. In thermal equilibrium where ${\bf Q}={\bf 0}$ and ${\bf k}=
{\bf 0}$, the long-range order implies a nonzero order parameter $\langle\psi
\rangle =\eta$ so that $\xi_1$ is infinite and $G({\bf r},{\bf r}^\prime)$
does not decay to zero but only to the absolute square of the order parameter
$\eta^2$. In the previous theory of Ref.\ \onlinecite{HD2} the heat flow is 
assumed to be metastable where dissipation by creation of vortices was 
neglected. Thus, in this theory long-range order is preserved also for 
nonzero $Q$ so that $\langle\psi({\bf r})\rangle = \eta\, e^{i{\bf kr}}$ is 
again nonzero, $\xi_1$ is infinite, and the Green's function does not decay 
to zero. 

However, in the present theory vortices are included implicitly so that 
$\langle\psi({\bf r})\rangle=0$, $\xi_1$ is finite, and $G({\bf r},
{\bf r}^\prime)$ decays to zero eventually for large distances. The second
correlation length $\xi_1$ represents a {\it dephasing length} of the 
order-parameter field $\psi({\bf r})$ which is equal to the average distance 
between the vortices. An approximate formula for $\xi_1$ can be extracted 
from the integral representation (\ref{Z28}) of the Green's function in the 
Appendix. In the superfluid region for $\Delta T$ sufficiently below 
$\Delta T_\lambda(Q)$ the exponential factor $\exp\{-\alpha\bar r_1 - 
\alpha^3 s\}$ as a function of $\alpha$ has a sharp maximum located at
\begin{equation}
\alpha_0 = (-\zeta)^{3/2} / [3^{1/2} (-\bar r_1)] 
\label{G03}
\end{equation}
where $\bar r_1$ and $\zeta=s^{-1/3} \bar r_1$ are negative. Thus, in the 
second line of (\ref{Z28}) we may replace $\alpha$ approximately by 
$\alpha_0$ so that the integral can be evaluated by (\ref{Z29}). As a result 
we obtain the Green's function
\begin{equation}
G({\bf r},{\bf r}^\prime) \approx n_{\rm s} \,\exp\bigl\{ i{\bf k}({\bf r}-
{\bf r}^\prime) - ({\bf r}-{\bf r}^\prime)^2/2\xi_1^2 \bigr\}  \ , 
\label{G04}
\end{equation}
where 
\begin{equation}
{\bf k} = \alpha_0 (F/8\gamma w^\prime)\, {\bf b} 
\label{G05}
\end{equation}
is the wave vector and 
\begin{equation}
\xi_1 = (2\alpha_0)^{1/2}
\label{G06}
\end{equation}
is the dephasing length. Clearly, Eq.\ (\ref{G04}) implies the asymptotic 
decay formula (\ref{G02}) where $\eta^2=n_{\rm s}$. Inserting (\ref{G03})
into (\ref{G05}) we find that ${\bf k}$ is exactly the average wave vector 
(\ref{Z41}) which we have calculated in the Appendix. By using $\bar r_1=
\xi^{-2} \rho_1$ and $\rho_1=\sigma^{1/3}\zeta$ we find that the dephasing 
length (\ref{G06}) can be written in the form 
\begin{equation}
\xi_1 = \xi \, \bigl[ {\textstyle{4\over 3}} (-\zeta) \bigr]^{1/4} 
\sigma^{-1/6}  \ . 
\label{G07}
\end{equation}
This formula is suitable for a numerical evaluation of $\xi_1$ because 
$\zeta$ and $\sigma$ are two of the seven dimensionless parameters, which are
determined in Sec.\ \ref{S04}. In Fig.\ \ref{F08} the dephasing length 
$\xi_1$ is shown as dashed line. We note that $\xi_1$ is defined only in the
superfluid region for $\Delta T\lesssim \Delta T_\lambda(Q)$. It turns out
that $\xi_1$ is always larger than $\xi$. While the correlation length $\xi$
decreases with increasing distance from criticality, the dephasing length
$\xi_1$ increases. 

By using the asymptotic formulas of Sec.\ \ref{S05}.B an asymptotic formula
for $\xi_1$ can be derived. Eliminating the amplitude $A$ from (\ref{D10}) 
and (\ref{D11}) we obtain
\begin{equation}
\sigma^{-1/6} \approx (-\zeta)^{1/2} {A_d\over 8u[\tau]} \ {g_0 k_{\rm B}
T_\lambda \over Q\,\xi^{d-1} }  \ . 
\label{G08}
\end{equation}
Consequently, from (\ref{G07}) we obtain
\begin{equation}
\xi_1 \approx \bigl[{\textstyle{4\over 3}} (-\zeta)^3 \bigr]^{1/4} 
{A_d\over 8u[\tau]} \ {g_0 k_{\rm B} T_\lambda \over Q\,\xi^{d-2} } \ . 
\label{G09}
\end{equation}
Since $\zeta$ is nearly constant, the leading dependences on $\Delta T$ and
$Q$ are governed by 
\begin{equation}
\xi_1 \sim Q^{-1} \xi^{-(d-2)} \sim Q^{-1} (-\Delta T)^{(d-2)\nu}  \ . 
\label{G10}
\end{equation}
This asymptotic formula clearly indicates that the dephasing length $\xi_1$ 
diverges in the limit $Q\rightarrow 0$. Thus, in thermal equilibrium where 
$Q=0$ the second correlation length $\xi_1$ is infinite as expected, which 
means that in this case vortices are not present. 

In the Appendix we have calculated the Green's function $G({\bf r},
{\bf r}^\prime)$ and the related quantities $n_{\rm s}$ and $J_{\rm s}$
approximately by linearizing the parameters $r_1$ and $\Delta r$ locally 
with respect to the space variable ${\bf r}$ according to (\ref{Z12}) and 
(\ref{Z13}). Since $r_1$ and $\Delta r$ are related to the temperature 
profiles $\Delta T(z)=T(z)-T_\lambda(z)$ and $T(z)$, respectively, and 
since $T_\lambda(z)$ is always a linear function in $z$, the validity of
the approximation is proven if the curvature $\partial_z^2 T$ of the 
temperature profile is negligible compared to the gradient $\partial_z T$.
For this purpose we define
\begin{equation}
\xi_2 = \vert \partial_z T \vert\, / \,\vert \partial_z^2 T \vert 
\label{G11}
\end{equation}
which may be viewed as the characteristic length scale for the variations of
the temperature profile $T(z)$ with respect to the space coordinate $z$. In
Fig.\ \ref{F08} the characteristic length $\xi_2$ is shown as dotted line. 
The curvature of $T(z)$ is negligible if $\xi_2$ is considerably larger than
the intrinsic correlation lengths $\xi$ and $\xi_1$ of the helium. 
In Fig.\ \ref{F08} the dotted line is considerably above the solid line and
the dashed line for temperatures $\Delta T$ sufficiently far away from 
$\Delta T_\lambda(Q)$. Consequently, the criterion for the validity of our 
approximation is satisfied in the normal-fluid region and in the superfluid 
region sufficiently far away from the superfluid-normal-fluid interface. 
Close to the interface all three correlation lengths $\xi$, $\xi_1$, and
$\xi_2$ have the same order of magnitude so that here the curvature of $T(z)$
is important and our approximation is strictly speaking invalid. However,
since nothing unusual is observed here, we believe that our theory yields 
reasonable and reliable interpolations for the physical quantities close to
the interface. 

The dotted line in Fig.\ \ref{F08} represents the characteristic length 
$\xi_2$ for a heat flow in zero gravity. Similar results for $\xi_2$ are
obtained if we insert the temperature profile of a vertical heat flow in
gravity. Deviations are expected for temperatures $\Delta T\lesssim \Delta 
T_1(Q)$. In gravity we find a somewhat smaller $\xi_2$, except for the 
downwards heat flow if $\Delta T$ is close to $\Delta T_1(Q)$. In the latter 
case $\xi_2$ diverges in the limit $\Delta T \rightarrow \Delta T_1(Q)$ which 
means that for the SOC state the approximation is exactly valid. Eventually, 
it turns out that the criterion for the validity of our approximation for 
calculating $G({\bf r},{\bf r}^\prime)$, $n_{\rm s}$, and $J_{\rm s}$ is not 
affected significantly by gravity.

\section{Entropy and specific heat} \label{S08}
In model {\it F\,} the field variable $m({\bf r},t)$ represents a fluctuating 
entropy density divided by $k_{\rm B}$. For this reason the local entropy 
density is defined by 
\begin{equation}
S = S_0 + k_{\rm B} \langle m \rangle 
\label{H01}
\end{equation}
where $S_0$ is a constant. Thus, the average $\langle m\rangle$ must be 
evaluated. From the free energy functional (\ref{A03}) we derive
\begin{equation}
\Bigl\langle {\delta H \over \delta m} \Bigr\rangle \ =\ \chi_0^{-1} 
\langle m\rangle + \gamma_0 \langle \vert \psi \vert^2 \rangle - h_0  \ . 
\label{H02}
\end{equation}
We insert this quantity into (\ref{B29}), replace $\langle\vert\psi\vert^2
\rangle =n_{\rm s}$, and obtain 
\begin{equation}
r_0(z) = \tau_0(z) + 2\chi_0\gamma_0^2 n_{\rm s} + 2\gamma_0 \langle m
\rangle  \ . 
\label{H03}
\end{equation}
From (\ref{B31}) we obtain $n_{\rm s}=(4u_0)^{-1} [r_1(z)-r_0(z)]$. We 
resolve (\ref{H03}) with respect to $\langle m\rangle$ and then obtain the
entropy
\begin{eqnarray}
S &=& S_0 + k_{\rm B} (2\gamma_0)^{-1} \Bigl\{ -\tau_0(z) + r_0(z) 
\nonumber\\ 
&&\hskip2cm + {\chi_0\gamma_0^2 \over 2u_0} [r_0(z)-r_1(z)] \Bigr\}  \ . 
\label{H04}
\end{eqnarray}
Within our approximation in Secs.\ \ref{S02} and \ref{S03} we find $r_0(z)=
r_1(z)=0$ at the critical point $(\Delta T,Q,g)=(0,0,0)$. Hence
\begin{equation}
S_\lambda = S_0 - k_{\rm B} (2\gamma_0)^{-1} \tau_0(z) 
\label{H05}
\end{equation}
is the entropy density at $T=T_\lambda$ in thermal equilibrium and zero 
gravity. We define the entropy difference $\Delta S=S-S_\lambda$ as the
deviation from the entropy at criticality $S_\lambda$. Then, from (\ref{H04})
and (\ref{H05}) we obtain 
\begin{equation}
\Delta S = k_{\rm B} {r_0(z) \over 2\gamma_0} \Bigl\{ 1 + {\chi_0\gamma_0^2
\over 2u_0} \Bigl[ 1 - {r_1(z) \over r_0(z)} \Bigr] \Bigr\} \ . 
\label{H06}
\end{equation}
While $S_\lambda$ is a constant, $\Delta S=\Delta S(\Delta T,Q)$ is strongly
influenced by the critical fluctuations near the superfluid transition. For
this reason, we must renormalize the entropy $\Delta S$ and apply the RG
theory. In analogy to the field variable $m({\bf r},t)$ the entropy is 
renormalized by \cite{D1}
\begin{equation}
\Delta S = (\chi_0 Z_m)^{1/2} \Delta S_{\rm ren}  \ . 
\label{H07}
\end{equation}
In (\ref{H06}) we replace the bare model-{\it F\,} parameters by the 
renormalized counterparts by using (\ref{C01}), (\ref{C05}), and (\ref{C06}).
As a result we obtain the renormalized entropy
\begin{equation}
\Delta S_{\rm ren} = k_{\rm B} \Bigl( {A_{\rm d} \over \mu^\epsilon} 
\Bigr)^{1/2} {r(z) \over 2\gamma} \Bigl\{ 1 + {\gamma^2 \over 2u} 
\Bigl[ 1 - {r_1(z) \over r(z)} \Bigr] \Bigr\} \ . 
\label{H08}
\end{equation}
Within the Hartree approximation all $Z$ factors cancel, where an expansion 
with respect to the coupling parameters is not necessary. By resolving 
(\ref{C25}) with respect to $\chi_0Z_m$ we obtain the renormalization factor
of (\ref{H07}) as 
\begin{equation}
(\chi_0 Z_m)^{1/2} = (A_d \mu^d)^{1/2}\, [ 2\tau\gamma]^{-1} \ . 
\label{H09}
\end{equation}
We replace $r(z)$ and $r_1(z)$ by the dimensionless variables $\rho$ and 
$\rho_1$ according to (\ref{C11}) and (\ref{C13}). We apply the RG theory
to (\ref{H07})-(\ref{H09}) by replacing $\mu\rightarrow \mu l=\xi^{-1}$,
$u\rightarrow u[\tau]$, $\gamma\rightarrow \gamma[\tau]$, etc., so that
now all dimensionless coupling parameters depend on the RG flow parameter 
$\tau$. Combining the resulting three equations together we eventually obtain 
the entropy difference
\begin{equation}
\Delta S = k_{\rm B} {A_{\rm d} \over 4\tau\, \xi^d} \,\rho\, \Bigl\{ 
{1\over\gamma[\tau]^2} + {1 \over 2u[\tau]} \Bigl[ 1 - {\rho_1\over \rho} 
\Bigr] \Bigr\} \ . 
\label{H10}
\end{equation}
Eq.\ (\ref{H10}) is the final formula which can be used for numerical
evaluation of the entropy density $S=S_\lambda + \Delta S$. There are no 
adjustable parameters present in this formula. The needed dimensionless
parameters $\tau$, $\rho$, and $\rho_1$ and the correlation length $\xi=
\xi(\tau)$ were determined in Sec.\ \ref{S04}. We note that Eq.\ (\ref{H10})
was derived within the Hartree approximation. It is not restricted to the 
physical situation considered in this paper where the helium is influenced 
by a heat current and gravity. The formula can be applied also to other 
physical situations as e.g.\ rotating helium whenever the Hartree 
approximation is used. 

The entropy $\Delta S$ is a static quantity so that nonasymptotic effects
of the dynamic RG theory are very small. For this reason, close to 
criticality Eq.\ (\ref{H10}) can be evaluated asymptotically by using the
asymptotic formulas $\xi=\xi_0 \tau^{-\nu}$, $u[\tau]\approx u^*=0.0362$,
and
\begin{equation}
{1\over \gamma[\tau]^2} \ =\ {4\nu\over \alpha} (1-b\,\tau^\alpha)  \ , 
\label{H11}
\end{equation}
where $\nu=0.671$ and $\alpha=2-d\nu=-0.013$ (see Refs.\ \onlinecite{D1} and
\onlinecite{SD}). Eq.\ (\ref{H11}) is obtained by solving the RG equation
for $\gamma[\tau]$ asymptotically. Since $\gamma[\tau]$ is a known function
\cite{D1}, $b$ is a known constant (which actually is close to unity). Now, 
inserting the asymptotic formulas into (\ref{H10}) we eventually obtain the 
entropy density 
\begin{equation}
S = S_\lambda + t\,\bigl\{ B + \tilde A\,[\,(4\nu / \alpha) + E[u^*]\,]
\tau^{-\alpha} \bigr\} 
\label{H12}
\end{equation}
where
\begin{eqnarray}
t&=& \tau\,\rho = \Delta T/T_\lambda \ , \label{H13} \\
E[u] &=& (2u)^{-1} [1-\rho_1/\rho] \ , \label{H14} \\
\tilde A &=& k_{\rm B} A_d / 4 \,\xi_0^d \ , \label{H15} \\
B &=& \tilde A (-4\nu/\alpha) b \ . \label{H16}
\end{eqnarray}

The specific heat $C$ is obtained by differentiation of the entropy $S$ with 
respect to temperature according to 
\begin{equation}
C = T_\lambda \, {\partial S \over \partial T} = {\partial S \over 
\partial t}  \ . 
\label{H17}
\end{equation}
From our experience the best way to calculate the specific heat is first
to calculate the entropy $S$ by (\ref{H10}) or (\ref{H12}) and then to
determine the specific heat $C$ by numerical differentiation via (\ref{H17}). 
(The alternative way, to differentiate the bare entropy (\ref{H04}) first with
respect to $r_0(z)$ and then to apply the RG theory, is less reliable and 
yields artifacts, so that it should not be used.) While the constants 
$\tilde A$ and $B$ are given by (\ref{H15}) and (\ref{H16}), we can
alternatively determine these constants by fitting the specific heat in 
thermal equilibrium to the data of the newest experiment \cite{LS}, which
was performed in microgravity in space. In this way we obtain $\tilde A=2.22
\ \mbox{J/mol}\,\mbox{K}$ and $B=456\ \mbox{J/mol}\,\mbox{K}$. Since in the 
experiments the {\it molar} specific heat is measured, we have multiplied 
the constants by the molar volume $V_\lambda$, which for saturated vapor 
pressure is \cite{TA} $V_\lambda=27.38\ \mbox{cm}^3/\mbox{mol}$. 

\begin{figure}[t]
\vspace*{6.9cm}
\includegraphics{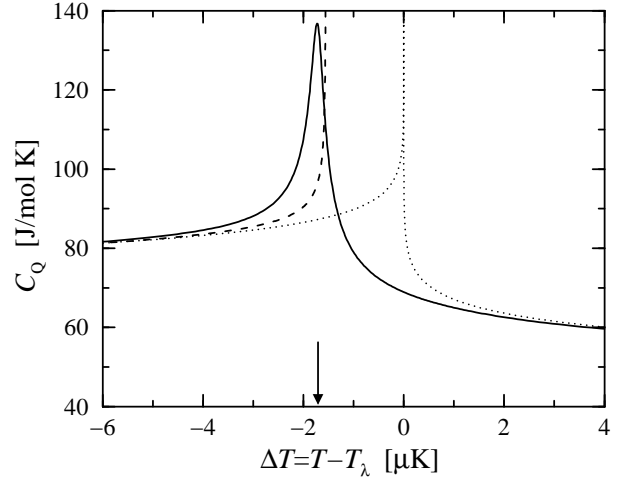}
\caption{The specific heat $C_Q(\Delta T,Q)$ as a function of $\Delta T$ for 
the constant heat current $Q=42.9\ \mu\mbox{W/cm}^2$. Our theoretical result 
obtained from (\ref{H12}) and (\ref{H17}) is shown as solid line. The dashed 
line represents $C_Q$ of the previous theory \protect\cite{HD3}, where 
vortices were neglected. For comparison we have plotted the specific heat in 
thermal equilibrium for $Q=0$ as dotted line. The arrow indicates $\Delta 
T_\lambda(Q)$.}
\label{F09}
\end{figure}

The specific heat $C$ depends on the thermodynamic variable which is kept
constant when performing the differentiation with respect to temperature
in (\ref{H17}). Since we consider liquid $^4$He in the presence of a constant
heat flow, the heat current $Q$ is the natural variable which should be
kept constant. For this reason we calculate $C_Q$ at constant $Q$. In
Fig.\ \ref{F09} $C_Q=C_Q(\Delta T,Q)$ is plotted as a function of $\Delta T$
for $Q=42.9\ \mu\mbox{W/cm}^2$, i.e.\ the same heat current as in 
Figs.\ \ref{F03} and \ref{F08}. Our theoretical result is shown as solid
line. For comparison, the specific heat in thermal equilibrium at $Q=0$ is
shown as dotted line. While in thermal equilibrium the specific heat is
singular at $\Delta T=0$, for nonzero $Q$ we find a smooth curve for 
$C_{\rm Q}$ which exhibits a strong maximum located at $\Delta T_\lambda(Q)$. 
We note that the temperature $T(z)$ is space dependent, so that the specific
heat $C_Q(z)$ is also space dependent and must be interpreted as a local
quantity.

Within the framework of the previous theory \cite{HD2}, the specific heats
$C_{v_{\rm s}}$ and $C_Q=C_{J_{\rm s}}$ were calculated and a similar 
formula like (\ref{H12}) was obtained for the entropy \cite{HD3}. In 
Fig.\ \ref{F09} the specific heat $C_Q$ of the previous theory is shown as 
dashed line. Since the heat flow is metastable below and unstable above
$\Delta T_\lambda(Q)$, the dashed line is defined only for $\Delta T<\Delta 
T_\lambda(Q)$. The specific heat $C_Q$ is enhanced by the nonzero $Q$ and 
diverges \cite{CG} at $\Delta T_\lambda(Q)$. Experiments to measure $C_Q$ at 
constant $Q$ are in progress \cite{HL}. An enhancement of the specific heat 
by a nonzero heat current was found just recently.

\begin{figure}[t]
\vspace*{6.9cm}
\includegraphics{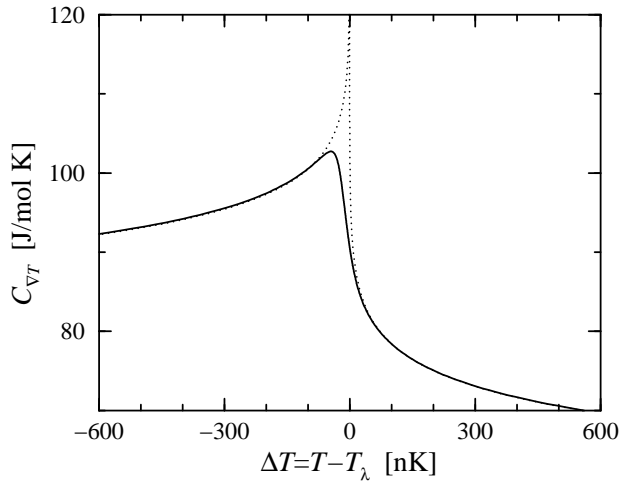}
\caption{The specific heat $C_{\bbox{\nabla}T}(\Delta T,Q)$ as a function 
of $\Delta T$ for the self organized critical (SOC) state where the 
temperature gradient is fixed by gravity according to $\partial_z T=
\partial_z T_\lambda=-1.273\ \mu \mbox{K/cm}$. The solid line represents our 
theoretical result obtained from (\ref{H12}) and (\ref{H17}). For comparison 
the specific heat in thermal equilibrium is shown as dotted line.}
\label{F10}
\end{figure}

The self organized critical (SOC) state represents an ideal system for 
measuring thermodynamic quantities like the specific heat because it is 
homogeneous in space over a large region. In this case the temperature 
gradient is fixed by gravity according to $\partial_z T=\partial_z T_\lambda=
-1.273\ \mu\mbox{K/cm}$. Thus, in the SOC state the specific heat 
$C_{\bbox{\nabla}T}$ at constant temperature gradient will be measured. We
calculate $C_{\bbox{\nabla}T}$ from (\ref{H17}) by inserting the entropy 
(\ref{H12}) and keeping $\partial_z T=\partial_z T_\lambda$ constant when
performing the numerical differentiation, while the heat current $Q$ is
varied appropriately. The result is shown in Fig.\ \ref{F10} as solid line.
For comparison, the equilibrium specific heat is shown as dotted line. 
Clearly, the nonzero temperature gradient $\partial_z T=\partial_z T_\lambda$
of the SOC state induced by gravity implies a rounding of the critical 
singularity. The solid line is smooth and exhibits a maximum at $\Delta 
T_g=-45\ \mbox{nK}$. While in Fig.\ \ref{F09} the maximum of $C_Q$
is very strong and enhanced, here in Fig.\ \ref{F10} the maximum of 
$C_{\bbox{\nabla}T}$ is moderate and just represents a smooth and round 
interpolation of the equilibrium specific heat near criticality. The 
temperature scale in Fig.\ \ref{F10} indicates that nano-Kelvin resolution is 
sufficient for a measurement of $C_{\bbox{\nabla}T}$. Thus, the rounded 
temperature dependence of $C_{\bbox{\nabla}T}$ should be accessible by
present-day experiments. Since gravity is needed for the realization of the 
SOC state, the experiment should be performed on earth.

\section{Vortices and mutual friction} \label{S09}
While our theory does not include vortices explicitly, here we present some
arguments, which support that our theory includes the effect of vortices 
implicitly. In Sec.\ \ref{S05} we have calculated a finite thermal
conductivity $\lambda_{\rm T}$ for $\Delta T\lesssim\Delta T_\lambda(Q)$ 
which implies dissipation of the heat current $Q$ in the superfluid state. 
Since the heat is transported convectively by the superfluid-normal-fluid
counterflow, a superfluid current with a velocity ${\bf v}_{\rm s}$ is 
induced in the opposite direction of the heat flow. The superfluid velocity 
${\bf v}_{\rm s}$ is related to the phase of the order parameter field
$\psi({\bf r},t)$, so that the dissipation of the superfluid current is
necessarily related to the creation of vortices. Thus, since our theory 
describes dissipation, it must include vortices in some way. 

Gorter and Mellink \cite{GM} investigated mutual friction of the counterflow
experimentally and proposed the mutual-friction force density
\begin{equation}
f = A\, \rho_{\rm n} \rho_{\rm s} (v_{\rm s}-v_{\rm n})^3 
\label{K01}
\end{equation}
where $A$ is the so called Gorter-Mellink coefficient which may be a function
of temperature but which should be independent of the velocities $v_{\rm s}$ 
and $v_{\rm n}$. The force density (\ref{K01}) was added to the hydrodynamic 
equations of the two-fluid model (see (8) and (9) in Ref.\ \onlinecite{GM}).
For a stationary counterflow the relation 
\begin{equation}
{\rho_{\rm n} \rho_{\rm s}\over \rho} \, s_\lambda \, \partial_z T=
A\, \rho_{\rm n} \rho_{\rm s} (v_{\rm s}-v_{\rm n})^3 
\label{K02}
\end{equation}
was found, where the pressure gradient is neglected close to criticality. 
This relation implies $\partial_z T \sim (v_{\rm s}-v_{\rm n})^3$ and means
that the temperature gradient induced by the mutual friction is proportional
to the third power of the counterflow velocity. By experimental and 
theoretical considerations Vinen \cite{V1} showed that the ansatz (\ref{K01}) 
is related to a turbulent superfluid flow which generates a tangle of vortices 
which imply the mutual friction. A statistical theory for the superfluid
turbulence in a homogeneous counterflow was developed by Schwarz \cite{S1}.
This latter theory supports (\ref{K01}) and (\ref{K02}). The Gorter-Mellink 
coefficient $A$ was calculated \cite{S1} for temperatures in the interval 
$1.2\ {\rm K}< T< 2.05\ {\rm K}$ and agreement with the experiments 
\cite{V1,BE} was found (see also the review by Tough \cite{T1}).

Now, here we show that the ansatz of Gorter and Mellink (\ref{K01}) can be 
derived from model {\it F\,} by our approximation. To do this, we resolve 
(\ref{K02}) with respect to $A$, obtain
\begin{equation}
A = {s_\lambda\over\rho} \, {\partial_z T\over (v_{\rm s}-v_{\rm n})^3} \ , 
\label{K03}
\end{equation}
and insert the results of our calculations. The coefficient $A$ must be 
proven to be independent of $Q$ and $v_{\rm s}-v_{\rm n}$. Near criticality
$v_{\rm n}$ can be neglected because it is much smaller than $v_{\rm s}$. 
The superfluid velocity $v_{\rm s}$ is related to the wave number $k$ of the
order parameter by $v_{\rm s}=\hbar k/m_4$. The entropy per mass $s_\lambda$
is related to the model-{\it F\,} parameter $g_0$ by \cite{TA} $s_\lambda=
(\hbar/m_4)(g_0/T_\lambda)$. The temperature gradient $\partial_z T$ is
related to the thermal conductivity $\lambda_{\rm T}$ by (\ref{E01}). Then,
from (\ref{K03}) we obtain the Gorter-Mellink coefficient
\begin{equation}
A = \Bigl( {m_4\over\hbar} \Bigr)^2 {g_0 \over \rho_\lambda T_\lambda} \, 
{Q\over \lambda_{\rm T} k^3} \ . 
\label{K04}
\end{equation}
In the previous theory \cite{HD2} the heat current $Q$ was calculated as a
function of the wave number $k$ where the heat flow is metastable and 
dissipation by creation of vortices is neglected. In the superfluid region 
sufficiently well below $\Delta T_\lambda(Q)$ the dissipation is small so
that the result for $Q=Q(\Delta T,k)$ of Ref.\ \onlinecite{HD2} may be used 
to replace $k$ by $Q$. In this region, $Q=Q(\Delta T,k)$ is approximately a 
linear function of $k$ given by 
\begin{equation}
Q = {g_0 k_{\rm B} T_\lambda \over \xi^{d-2}} \, A_d\, \Bigl( {1\over 
8 u[\tau]} + {1\over d} \Bigr) \, k  \ .
\label{K05}
\end{equation}
Consequently, we obtain
\begin{eqnarray}
A &=& \Bigl( {m_4\over\hbar} \Bigr)^2 {g_0 \over \rho_\lambda T_\lambda} \, 
{1\over \lambda_{\rm T} Q^2} \nonumber\\
&&\times\, \Bigl\{  {g_0 k_{\rm B} T_\lambda \over \xi^{d-2}} \, A_d\, 
\Bigl( {1\over 8 u[\tau]} + {1\over d} \Bigr) \Bigr\}^3 \ \ . 
\label{K06}
\end{eqnarray}
The last factor $\{\cdots\}$ does not depend explicitly on $Q$. It only 
depends on the RG flow parameter $\tau$ via $\xi=\xi_0\tau^{-\nu}$ where
$u[\tau]\approx u^*=0.0362$. Now, inserting the asymptotic formula (\ref{D13})
for the thermal conductivity $\lambda_{\rm T}$ we clearly see that the heat
current $Q$ cancels. Neglecting the one-loop contribution in the heat-current
formula (\ref{K05}) we obtain the Gorter-Mellink coefficient
\begin{equation}
A \approx \Bigl( {m_4\over \hbar} \Bigr)^2 \, {g_0\over \rho_\lambda} \,
{\sqrt{12}\over (-\zeta)^{3/2}} \, {4\gamma[\tau] w^\prime[\tau]\over F[\tau]}
\ \tau\, \xi^2 
\label{K07}
\end{equation}
which does not depend explicitly on $Q$. A weak indirect $Q$ dependence is
included in the dimensionless variable $\zeta$, which represents the 
nonasymptotic effects of the dynamic RG theory. 
In leading order we find the asymptotic formula
\begin{equation}
A \sim \tau\, \xi^2 \sim (-\Delta T)^{1-2\nu} 
\label{K08}
\end{equation}
This result proves that within the approximation of our theory the ansatz
of Gorter and Mellink is correct and can be derived from model {\it F}. 

\begin{figure}[t]
\vspace*{6.9cm}
\includegraphics{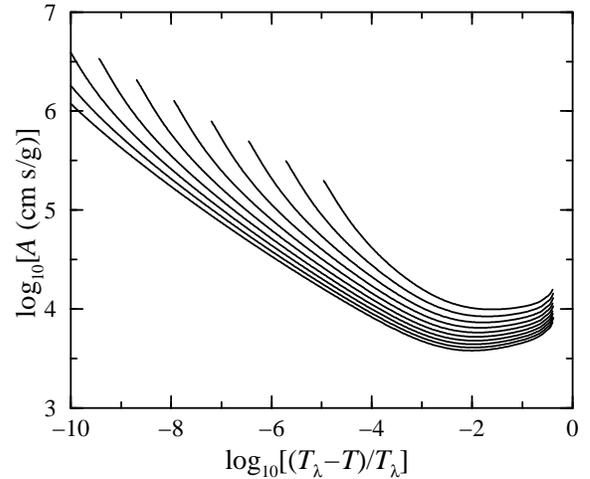}
\caption{The Gorter-Mellink coefficient $A$ obtained from (\ref{K06}) 
as a function of temperature for several heat currents between 
$Q=1\ \mbox{pW/cm}^2$ and $1\ \mbox{mW/cm}^2$ (curves from left to right, 
heat currents increase by a factor of $10$ respectively). The left ends of 
the curves are located close to the superfluid transition at 
$\Delta T_\lambda(Q)$.}
\label{F11}
\end{figure}

Eq.\ (\ref{K06}) can be used for an explicit calculation of the Gorter-Mellink
coefficient $A=A(\Delta T,Q)$ as a function of $\Delta T=T-T_\lambda$ and $Q$,
where our result for $\lambda_{\rm T}=\lambda_{\rm T}(\Delta T,Q)$ from 
Sec.\ \ref{S05} is inserted. While most constants and parameters are known, 
we additionally need $m_4/\hbar= 6320\ \mbox{s/cm}^2$ and the density 
$\rho_\lambda$ of the helium at $T_\lambda$, which at saturated vapor 
pressure is given by \cite{TA} $\rho_\lambda=N_{\rm A}m_4/V_\lambda=0.146
\ \mbox{g/cm}^3$. In Fig.\ \ref{F11} our theoretical prediction for $A$ is 
shown versus $\Delta T$ double logarithmically for several heat currents $Q$ 
which change by a factor of $10$ between each curve. The lowest and most left 
curve corresponds to the smallest heat current $Q=1\ \mbox{pW/cm}^2$, while 
the highest and most right curve corresponds to the largest heat current 
$Q=1\ \mbox{mW/cm}^2$. Since the helium is superfluid only for $\Delta T 
\lesssim \Delta T_\lambda(Q)$, all curves have an endpoint on the left-hand 
side. 

If the cancellation of $Q$ in (\ref{K06}) is perfect, then the curves in 
Fig.\ \ref{F11} would lie all on the same line. This, however, is not the
case. The nonasymptotic effects of the dynamic RG theory imply that $A(\Delta
T,Q)$ depends weakly on $Q$. While the heat current $Q$ is varied over $9$
decades, $A$ changes by a factor of $3$ to $10$ or by $0.5$ to $1.0$ decades.
Consequently, we approximately find
\begin{equation}
A \sim Q^y \sim (v_{\rm s}-v_{\rm n})^y 
\label{K09}
\end{equation}
with an exponent $y$ between $0.05$ and $0.1$. This result slightly modifies
the mutual friction force (\ref{K01}) of Gorter and Mellink into
\begin{equation}
f \sim (v_{\rm s}-v_{\rm n})^\phi
\label{K10}
\end{equation}
with the exponent $\phi=3+y$ between $3.05$ and $3.1$. 

The Gorter-Mellink coefficient $A$ was measured as a function of temperature
by Vinen \cite{V1}. The value $A\approx 200\ \mbox{cm}\,\mbox{s/g}$ was 
obtained for $\Delta T\approx -0.1\ \mbox{K}$, which can be extrapolated to 
$A\sim 600\ \mbox{cm}\,\mbox{s/g}$ for $\Delta T\sim -0.01\ \mbox{K}$. Similar 
values for $A$ were obtained also in later experiments, which are reviewed in 
Ref.\ \onlinecite{T1}. On the other hand, the lowest values, that our theory 
predicts, are about $A\sim 10^4\ \mbox{cm}\,\mbox{s/g}$ for $\Delta T\sim 
-0.01\ \mbox{K}$. Thus, our theoretical prediction for $A$ is about one or two 
decades larger than the experimental values. Furthermore, $A$ can be extracted 
from the experimental data of Baddar et al.\ \cite{BA} by inserting the 
power-law formula (\ref{D15}) for the thermal conductivity $\lambda_{\rm T}$ 
into (\ref{K06}). Again, our theoretical prediction is about a factor of $20$ 
larger than the experimentally observed values. Thus, we conclude that our 
theory has a tendency to overestimate the magnitude of the dissipation and 
mutual friction due to creation of vortices. On the other hand, the 
experimental data of Ref.\ \onlinecite{BA} imply the exponent $\phi=3.53$ for 
the mutual friction force (\ref{K10}), which is considerably larger than the 
exponent proposed by Gorter and Mellink \cite{GM} and obtained from our 
theory. 

The discrepancies are possibly due the number of vortices in the helium in
the presence of a homogeneous heat flow $Q$, because the number of vortices
appears to be an unknown and uncontrolled quantity in theory and experiment.
In rotating helium mutual friction can by studied in a much more controlled
fashion \cite{HV}, because in this case the number of vortices is related to 
the rotation frequency. Here the mutual friction is described by the Vinen
coefficients \cite{HV} $B$ and $B^\prime$. These coefficients have been 
measured experimentally (see Ref.\ \onlinecite{BD} for review) and also
calculated within model {\it F\,} in renormalized mean-field theory \cite{H3}
by considering the motion of a single vortex line. We have applied our theory 
(the Hartree approximation combined with the RG theory) also to rotating 
helium and find the coefficients \cite{H4} $B=(4m_4/\hbar)\Gamma^\prime[\tau]$ 
and $2-B^\prime=(4m_4/\hbar)\Gamma^{\prime\prime}[\tau]$. While these results 
are indeed simple, they have the same order of magnitude than the coefficients 
$B$ and $B^\prime$ of the previous theory \cite{H3} and also of the 
experiments \cite{BD}. The agreement is better for $B$ than for $B^\prime$. 
Thus, we conclude that our present theory indeed includes the effects of 
vortices. However, since a simple approximation is applied, the Hartree 
approximation (see Fig.\ \ref{F02}), discrepancies are expected. Nevertheless, 
for the Vinen coefficients $B$ and $B^\prime$ the discrepancies are much 
smaller than for the Gorter-Mellink coefficient $A$.

\section{Conclusions} \label{S10}
We have presented a renormalization-group (RG) theory based on model {\it F\,}
for liquid $^4$He near the superfluid transition in the presence of a heat
current $Q$ and gravity. The fundamental concept is a self-consistent 
approximation, which in quantum many-particle theory is known as the Hartree
approximation, combined with the RG theory. While above $T_\lambda$ the 
previous theory of Ref.\ \onlinecite{HD1} is recovered for a heat flow in 
normal fluid $^4$He, below $T_\lambda$ in the superfluid state our theory 
predicts dissipation of the heat current and mutual friction of the related 
superfluid-normal-fluid counterflow. 

We derived the ansatz of Gorter and Mellink \cite{GM} for the mutual friction
force and found several indications that our approach includes vortices
indirectly. However, our approach appears to overestimate the magnitude of 
the dissipation by vortex creation considerably compared to the experimental
observations. This discrepancy is probably due to the number of vortices in
the superfluid helium which appears to be an uncertain and very sensitive
quantity strongly influenced by the kind of the approximation and also by
the experimental conditions. Further theoretical and experimental work is
necessary to clarify the discrepancies. 

Besides the correlation length $\xi=\xi_0 \tau^{-\nu}$, which is the 
conventional length scale of the RG theory, in the superfluid state we find
a second characteristic length $\xi_1$ which describes the decay of the
correlations by dephasing of the order parameter field $\psi({\bf r})$ caused
by vortices. The dephasing length $\xi_1$ may be viewed as the average 
distance between the vortices. It is much larger than $\xi$ (see 
Fig.\ \ref{F08}) and approximately given by $\xi_1\approx a\, Q^{-1} 
\tau^{+\nu}$ where $\tau=-2\Delta T/T_\lambda$ and $a\approx 
1.0\ \mbox{mW/cm}$. Thus, for superfluid $^4$He in confined geometries we
expect novel effects when the dephasing length $\xi_1$ is as large as the
geometry and hits the boundary walls. Future theoretical and experimental
investigations should study the influence of vortices in superfluid $^4$He
on finite-size effects and on boundary effects. Unexpected results for the
thermal conductivity in confined geometries and for the Kapitza resistance
may possibly be found, which are caused by the second characteristic length
$\xi_1$.

\acknowledgements
I would like to thank Professors\ G.\ Ahlers, V.\ Dohm, R.\ Duncan, and 
H.\ Meyer for valuable discussions. Furthermore I would like to thank 
Drs.\ M.\ Lee, U.\ Israelsson, and D.\ Strayer for the invitation to the 1998 
NASA/JPL workshop on fundamental physics in microgravity, held in Oxnard,
CA, where many question about the relation between theory and experiment 
were discussed and clarified.

\appendix\section*{Integral representation for the Green's function}
In Sec.\ \ref{S03} and in Ref.\ \onlinecite{HD1} we have evaluated $n_{\rm s}=
\langle\vert\psi\vert^2\rangle$ and ${\bf J}_{\rm s}=\langle{\rm Im}[\psi^*
\bbox{\nabla}\psi]\rangle$ explicitly, which eventually are expressed in
terms of an integral by (\ref{C27}) and (\ref{C28}). While for our purposes 
we only need $n_{\rm s}$ and ${\bf J}_{\rm s}$, it is also possible to 
evaluate the complete matrix Green's function $G$ explicitly. This Green's
function is defined by (\ref{B18}). First of all, we need the inverse of the
operator matrix (\ref{B17}) which is given by 
\begin{equation}
\pmatrix{K_{\alpha\beta}^{-1}} = \pmatrix{ 0 & 2(L^+)^{-1} \cr
2 L^{-1} & 4\Gamma^\prime L^{-1} (L^+)^{-1} \cr } \ . 
\label{Z01}
\end{equation}
Since we intend to evaluate the renormalized Green's function, we use 
(\ref{C04}) for the operator $L$, where all model-{\it F\,} parameters are
replaced by the renormalized parameters. 

Here we consider the Green's function $\langle\psi\psi^*\rangle$ which is 
given by the lower right element $G_{22}$ of the matrix Green's function $G$. 
From (\ref{B18}) and (\ref{Z01}) we obtain
\begin{equation}
\langle\psi({\bf r},t)\psi^*({\bf r}^\prime,t^\prime)\rangle = 4\Gamma^\prime
\, L^{-1} (L^+)^{-1} \,\delta({\bf r}-{\bf r}^\prime) \,\delta(t-t^\prime) \ .
\label{Z02}
\end{equation}
We represent the inverse operators $L^{-1}$ and $(L^+)^{-1}$ as integrals of
exponential functions so that the Green's function is rewritten as
\begin{eqnarray}
\langle\psi({\bf r},t)\psi^*({\bf r}^\prime,t^\prime)\rangle &=& 
4\Gamma^\prime \int_0^\infty d\alpha \int_0^\infty d\beta \ e^{-\alpha L} 
\, e^{-\beta L^+}  \nonumber\\
&&\hspace{0.5cm} \times \,\delta({\bf r}-{\bf r}^\prime) \,\delta(t-t^\prime) 
\ .
\label{Z03}
\end{eqnarray}
We decompose the operators in the exponentials into time-dependent and
space-dependent parts according to
\begin{eqnarray}
-\alpha L&=&-\alpha \partial_t + \alpha A \ ,  \label{Z04}  \\
-\beta L^+&=&+\beta \partial_t + \beta B \ ,  \label{Z05}
\end{eqnarray}
where
\begin{eqnarray}
A&=&- \bigl\{ \Gamma [r_1 - \bbox{\nabla}^2] - i(g/2\gamma) \Delta r \bigr\}
\ ,  \label{Z06}  \\
B&=&- \bigl\{ \Gamma^* [r_1 - \bbox{\nabla}^2] + i(g/2\gamma) \Delta r \bigr\}
\ .  \label{Z07}
\end{eqnarray}
Since $\partial_t$ commutes with the space-dependent operators $A$ and $B$,
we obtain
\begin{eqnarray}
\langle\psi({\bf r},t)\psi^*({\bf r}^\prime,t^\prime)\rangle &=& 
4\Gamma^\prime \int_0^\infty d\alpha \int_0^\infty d\beta  \nonumber\\
&&\hspace{0.5cm}\times e^{\alpha A} \, e^{\beta B}  
\,\delta({\bf r}-{\bf r}^\prime)  \nonumber\\
&&\hspace{0.5cm}\times e^{(-\alpha +\beta)\partial_t} \,\delta(t-t^\prime) \ .
\label{Z08}
\end{eqnarray}
The time-dependent factor is evaluated by the Taylor series according to
\begin{equation}
e^{(-\alpha + \beta)\partial_t}  \,\delta(t-t^\prime) = \delta(-\alpha
+\beta+t-t^\prime) \ .
\label{Z09}
\end{equation}
The space-dependent factor is more complicated and can be evaluated by using
the formula
\begin{eqnarray}
e^A \, e^B &=& \exp\bigl\{ A+B + {\textstyle {1\over 2}} [A,B] \nonumber\\
&&\hspace{0.8cm} + {\textstyle {1\over 12}} \bigl( [A,[A,B]]-[B,[A,B]] \bigr) 
\nonumber\\ 
&&\hspace{0.8cm} - {\textstyle {1\over 24}} [B,[A,[A,B]]] + \cdots \bigr\} \ . 
\label{Z10}
\end{eqnarray}

\subsection{The equal time Green's function}
For simplicity we set $t^\prime=t$, and from now on we consider only the
equal time Green's function. The integral over $\beta$ can be evaluated 
easily by the delta function (\ref{Z09}) so that we obtain
\begin{eqnarray}
G({\bf r},{\bf r}^\prime) &=& \langle\psi({\bf r},t)\psi^*({\bf r}^\prime,t)
\rangle \nonumber\\
&=&4\Gamma^\prime \int_0^\infty d\alpha \ e^{\alpha A} \, e^{\alpha B}  
\,\delta({\bf r}-{\bf r}^\prime)  \ .  
\label{Z11}
\end{eqnarray}
For a successful calculation of the space-dependent integrand the series in 
the exponential on the right-hand side of (\ref{Z10}) must be finite, which 
means that only a finite number of the commutators may be nonzero. For this
reason, we assume $r_1$ and $\Delta r$ to be linear functions of the space 
coordinate $\bf r$ given by
\begin{eqnarray}
r_1 &=& a_1 + {\bf b}_1 {\bf r} \ ,  \label{Z12} \\
\Delta r &=& a + {\bf b} {\bf r} \ ,  \label{Z13}
\end{eqnarray}
where $a_1$ and $a$ are constants and ${\bf b}_1$ and ${\bf b}$ are constant
vectors. In general, $r_1$ and $\Delta r$ are nonlinear functions of $\bf r$.
In this case $r_1$ and $\Delta r$ must be linearized locally so that 
Eqs.\ (\ref{Z12}) and (\ref{Z13}) are taken as an approximation where 
${\bf b}_1$ and ${\bf b}$ are the respective gradients at the space point 
${\bf R} = {1\over 2}({\bf r}+{\bf r}^\prime)$. In the main text of this 
paper we assume $r_1(z)$ and $\Delta r(z)$ to be functions of the coordinate
$z$ only so that the gradients ${\bf b}_1=r_1^\prime {\bf e}_z$ and
${\bf b}=\Delta r^\prime {\bf e}_z$ are vectors in $z$ direction. However,
this latter assumption is not necessary here. 

Now, Eqs.\ (\ref{Z12}) and (\ref{Z13}) imply the commutators $[r_1,
\bbox{\nabla}^2]=-2\,{\bf b}_1 \bbox{\nabla}$ and $[\Delta r,\bbox{\nabla}^2] 
=-2\,{\bf b} \bbox{\nabla}$. Consequent\-ly, we find 
\begin{eqnarray}
{[A,B]} &=& - i (g\Gamma^\prime/\gamma)\, 2\, {\bf b}\bbox{\nabla} \ ,
\label{Z14} \\ 
{[A,[A,B]\,]} &=& -2i (g\Gamma^\prime/\gamma) \bigl\{ \Gamma\, {\bf b}_1 
{\bf b} - i (g/2\gamma) {\bf b}^2 \bigr\} \ , \label{Z15} \\
{[B,[A,B]\,]} &=& -2i (g\Gamma^\prime/\gamma) \bigl\{ \Gamma^* {\bf b}_1 
{\bf b} + i (g/2\gamma) {\bf b}^2 \bigr\} \ . \label{Z16} 
\end{eqnarray}
The double commutators (\ref{Z15}) and (\ref{Z16}) are {\it c} numbers because
${\bf b}_1$ and ${\bf b}$ are assumed to be constant. Hence, all higher-order
commutators are zero so that the series in the exponential on the right-hand 
side of (\ref{Z10}) is finite. Eventually, inserting the commutators 
(\ref{Z14})-(\ref{Z16}) into (\ref{Z10}) we obtain 
\begin{eqnarray}
e^{\alpha A} \, e^{\alpha B} &=& \exp \Bigl\{ - 2\Gamma^\prime\alpha
[r_1 - \bbox{\nabla}^2] - i\, ( 2\Gamma^\prime\alpha )^2 {g\over 4\gamma
\Gamma^\prime} {\bf b} \bbox{\nabla} \nonumber\\
&& \hspace{-1.5cm}
+ {1\over 12} ( 2\Gamma^\prime\alpha )^3 \Bigl[ 2 {\Gamma^{\prime\prime}
\over \Gamma^\prime} \Bigl( {g\over 4\gamma\Gamma^\prime} {\bf b} \Bigr) 
{\bf b}_1 - 4 \Bigl( {g\over 4\gamma \Gamma^\prime} {\bf b} \Bigr)^2 \Bigr] 
\Bigr\} \ . 
\label{Z17} 
\end{eqnarray}
This formula is exact if $r_1$ and $\Delta r$ are linear functions in $\bf r$
given by (\ref{Z12}) and (\ref{Z13}).

Next, we consider the operators $C=-2\Gamma^\prime r_1$ and $D=2\Gamma^\prime
(\bbox{\nabla}-i{\bf k})^2$ where ${\bf k}$ is a constant wave vector. We
find the commutators 
\begin{eqnarray}
{[C,D]} &=& (2\Gamma^\prime)^2 \, 2\, {\bf b}_1 (\bbox{\nabla}-i{\bf k}) \ , 
\label{Z18} \\ 
{[C,[C,D]\,]} &=& (2\Gamma^\prime)^3 \, 2\, {\bf b}_1^2 \ , \label{Z19} \\ 
{[D,[C,D]\,]} &=& 0 \ . \label{Z20} 
\end{eqnarray}
Since the double commutators are {\it c} numbers, again the higher-order 
commutators are zero. Thus, from (\ref{Z10}) we obtain the formula
\begin{eqnarray}
e^{\alpha C} \, e^{\alpha D} &=& \exp \bigl\{ - 2\Gamma^\prime\alpha
[r_1 - (\bbox{\nabla}-i{\bf k})^2] \nonumber\\
&&\hspace{-1cm} + (2\Gamma^\prime\alpha)^2 {\bf b}_1 (\bbox{\nabla}-i{\bf k}) 
+ {\textstyle {1\over 6}} (2\Gamma^\prime\alpha)^3 {\bf b}_1^2  \bigr\} \ .  
\label{Z21} 
\end{eqnarray}
We choose the wave vector 
\begin{equation}
{\bf k} = \Gamma^\prime \alpha \Bigl[ {g\over 4\gamma\Gamma^\prime} {\bf b}
- i {\bf b}_1 \Bigr] 
\label{Z22}
\end{equation}
and then rewrite (\ref{Z21}) as 
\begin{eqnarray}
e^{\alpha C} \, e^{\alpha D} &=& \exp \Bigl\{ - 2\Gamma^\prime\alpha
[r_1 - \bbox{\nabla}^2] - i\, ( 2\Gamma^\prime\alpha )^2 {g\over 4\gamma
\Gamma^\prime} {\bf b} \bbox{\nabla} \nonumber\\
&&- {1\over 12} ( 2\Gamma^\prime\alpha )^3 \Bigl[ {\bf b}_1^2 + 3 \Bigl( 
{g\over 4\gamma \Gamma^\prime} {\bf b} \Bigr)^2 \Bigr] \Bigr\} \ . 
\label{Z23} 
\end{eqnarray}
Now, we clearly see that the operators in the exponentials on the right-hand 
sides of (\ref{Z17}) and (\ref{Z23}) are equal up to an additive constant.
Thus, from comparison of (\ref{Z17}) and (\ref{Z23}) we obtain the relation
\begin{equation}
e^{\alpha A} \, e^{\alpha B} = e^{-(2\Gamma^\prime\alpha)^3 s} \cdot
e^{\alpha C} \, e^{\alpha D}
\label{Z24}
\end{equation}
where 
\begin{equation}
s = - {1\over 12} \Bigl[ {\bf b}_1^2 + 2 {w^{\prime\prime} \over w^\prime}
\Bigl( {F\over 4\gamma w^\prime} {\bf b} \Bigr) {\bf b}_1 - 
\Bigl( {F\over 4\gamma w^\prime} {\bf b} \Bigr)^2 \Bigr]  \ .  
\label{Z25}
\end{equation}
Here we have replaced $\Gamma^\prime$, $\Gamma^{\prime\prime}$, and $g$ by the 
dimensionless ratios $w^\prime = \Gamma^\prime/\lambda$, $w^{\prime\prime} = 
\Gamma^{\prime\prime}/\lambda$, and $F=g/\lambda$. The first factor on the 
right-hand side of (\ref{Z24}) is just a constant, while the other exponential
factors are operators. We insert the relation (\ref{Z24}) together with the
operators $C= -2\Gamma^\prime (a_1 + {\bf b}_1 {\bf r})$ and $D= 2
\Gamma^\prime (\bbox{\nabla} - i{\bf k})^2$ into (\ref{Z11}) and then obtain 
the Green's function 
\begin{eqnarray}
G({\bf r},{\bf r}^\prime) &=& 4\Gamma^\prime \int_0^\infty d\alpha 
\ e^{-2(\Gamma^\prime\alpha)^3 s} 
\, e^{-2\Gamma^\prime\alpha (a_1 + {\bf b}_1{\bf r})}  \nonumber\\
&&\hspace{1.5cm} \times \, e^{2\Gamma^\prime\alpha (\bbox{\nabla}-i{\bf k})^2}  
\,\delta({\bf r}-{\bf r}^\prime)  \ .  
\label{Z26}
\end{eqnarray}
We substitute $2\Gamma^\prime\alpha \rightarrow \alpha$ and evaluate the last 
two factors by using the Gaussian integral and the Taylor series:
\begin{eqnarray}
e^{\alpha (\bbox{\nabla}-i{\bf k})^2} \,\delta({\bf r}-{\bf r}^\prime) &=&
\nonumber\\
&&\hspace{-2,5cm} = (4\pi\alpha)^{-d/2} \int d^du \ e^{-(4\alpha)^{-1} 
{\bf u}^2 + {\bf u} (\bbox{\nabla}-i{\bf k})} \,\delta({\bf r}-{\bf r}^\prime)  
\nonumber\\ 
&&\hspace{-2.5cm} = (4\pi\alpha)^{-d/2} \int d^du \ e^{-(4\alpha)^{-1} 
{\bf u}^2 - i{\bf k} {\bf u} } \,\delta({\bf r}-{\bf r}^\prime+{\bf u})  
\nonumber\\ 
&&\hspace{-2.5cm} = (4\pi\alpha)^{-d/2} \ e^{i{\bf k}({\bf r}-{\bf r}^\prime)}
\, e^{-(4\alpha)^{-1} ({\bf r}-{\bf r}^\prime)^2 } 
\label{Z27}
\end{eqnarray}
Thus, inserting the wave vector (\ref{Z22}) with the proper subsitution for 
$\alpha$ and with $g/\Gamma^\prime$ replaced by $F/w^\prime$, for the 
Green's function we eventually obtain the formula 
\begin{eqnarray}
G({\bf r},{\bf r}^\prime) &=& {2\over (4\pi)^{d/2}} \int_0^\infty {d\alpha 
\over \alpha^{d/2}} \ \exp\{ -\alpha \bar r_1 - \alpha^3 s \}  \nonumber\\
&& \hspace{-1cm} \times \,\exp \Bigl\{ i\,\alpha\, 
{F\over 8\gamma w^\prime}\, {\bf b} ({\bf r}-{\bf r}^\prime) - 
{1\over 4\alpha} ({\bf r}-{\bf r}^\prime)^2 \Bigr\} \ . 
\label{Z28}
\end{eqnarray}
where $\bar r_1 = a_1 + {\textstyle{1\over 2}} {\bf b}_1 ({\bf r} + 
{\bf r}^\prime)$. This formula is a simple integral, there are no operators 
in the integrand any more. Clearly, the Green's function depends on the 
average coordinate ${\bf R}={1\over 2} ({\bf r}+{\bf r}^\prime)$ implicitly 
via $\bar r_1$ and explicitly on the relative coordinate $\Delta {\bf r} = 
{\bf r}-{\bf r}^\prime$. 

The integral (\ref{Z28}) has a similar but more general structure than the
integral ${\cal F}_\alpha(\zeta)$ in (\ref{C28}). Thus, it is possible to
obtain $n_{\rm s}$ and $J_{\rm s}$ from the Green's function (\ref{Z28}). 
We find
\begin{eqnarray}
n_{\rm s} &=& \langle \vert\psi\vert^2 \rangle = G({\bf r},{\bf r})
\nonumber\\
&=& {2\over (4\pi)^{d/2}} \int_0^\infty {d\alpha \over \alpha^{d/2}}
\ \exp\{-\alpha r_1 - \alpha^3 s\} 
\label{Z29}
\end{eqnarray}
and
\begin{eqnarray}
{\bf J}_{\rm s} &=& \langle {\rm Im} [\psi^* \bbox{\nabla}\psi ] \rangle = 
(2i)^{-1} [\bbox{\nabla}-\bbox{\nabla}^\prime] \, G({\bf r},{\bf r}^\prime)
\,\big\vert_{{\bf r}^\prime={\bf r}} \nonumber\\
&=& {F\over 4\gamma w^\prime} \, {{\bf b}\over (4\pi)^{d/2}} \int_0^\infty 
{d\alpha \over \alpha^{d/2-1}} \ \exp\{-\alpha r_1 - \alpha^3 s\}  \ . 
\nonumber\\
\label{Z30}
\end{eqnarray}
We identify ${\bf b}_1=r_1^\prime {\bf e}_z$ and ${\bf b}=\Delta r^\prime
{\bf e}_z$ and find that $s$ defined in (\ref{Z25}) is closely related to
$\sigma$ defined in (\ref{C29}). We substitute $\alpha=s^{-1/3}v$, identify
$\zeta=s^{-1/3} r_1$, and rewrite $n_{\rm s}$ and $J_{\rm s}$ as
\begin{eqnarray}
n_{\rm s} &=& {2\over (4\pi)^{d/2}} \, s^{(d-2)/6} \, {\cal F}_{(2-d)/2}
(\zeta) \ ,  \label{Z31} \\
J_{\rm s} &=& {F\over 4\gamma w^\prime} \, {\Delta r^\prime\over (4\pi)^{d/2}} 
\, s^{(d-4)/6} \, {\cal F}_{(4-d)/2}(\zeta) \ ,  \label{Z32}
\end{eqnarray}
where ${\bf J}_{\rm s}=J_{\rm s} {\bf e}_z$. By using the function 
$\Phi_\alpha(X)$ defined in (\ref{C27}) and identifying 
\begin{equation}
{1\over (4\pi)^{d/2}} = - {1\over\epsilon} \, A_d \, {1\over \Gamma(1-d/2)}
\label{Z33}
\end{equation}
which may be viewed as the definition of the factor $A_d$, the formulas for
$n_{\rm s}$ and $J_{\rm s}$ can be rewritten as 
\begin{eqnarray}
n_{\rm s} &=& - {2\over \epsilon} \, A_d \ \Phi_{-1+\epsilon/2}(X) \  
r_1^{1-\epsilon/2} \ ,  \label{Z34} \\
J_{\rm s} &=& {F\over 2w^\prime} \, {\Delta r^\prime \over 2\gamma} \,
{1\over \epsilon}\, A_d (1-\epsilon/2) \ \Phi_{\epsilon/2}(X) \  
r_1^{-\epsilon/2} \ ,  \label{Z35}
\end{eqnarray}
where $X=-s/r_1^3$ and $\epsilon=4-d$. Finally, replacing the renormalized 
couplings by the bare model-{\it F\,} parameters, we recover the formulas 
(\ref{B24}) and (\ref{B25}). Thus, we have derived the formulas of 
Sec.\ \ref{S03}.B for $n_{\rm s}$ and $J_{\rm s}$ once again. We note that 
the fundamental assumption here is the linear form of $r_1$ and $\Delta r$ 
in (\ref{Z12}) and (\ref{Z13}), which in general is an approximation. This
assumption implies the special structure of the integrals in (\ref{Z28}) and
in (\ref{C28}). 

\subsection{Fourier transformation of the Green's function
\break and physical interpretation}
The natural space variables of the equal time Green's function $G({\bf r},
{\bf r}^\prime)$ are the average coordinate ${\bf R}={1\over 2}({\bf r}+
{\bf r}^\prime)$ and the relative coordinate $\Delta {\bf r} = {\bf r} -
{\bf r}^\prime$. This fact is clearly seen in (\ref{Z28}). Thus, we may 
perform a Fourier transformation with respect to $\Delta {\bf r}$ and 
define the Fourier-transformed Green's function $\tilde G({\bf R},{\bf k})$
by 
\begin{equation}
G({\bf r},{\bf r}^\prime) = \int {d^dk\over (2\pi)^d} \ e^{i{\bf k}({\bf r}
-{\bf r}^\prime)} \ \tilde G({\bf R},{\bf k}) \ . 
\label{Z36}
\end{equation}
We apply this Fourier transformation to (\ref{Z28}) and then obtain
\begin{eqnarray}
\tilde G({\bf R},{\bf k}) &=& 2 \int_0^\infty d\alpha \ \exp\{ -\alpha \bar 
r_1 - \alpha^3 s \} \nonumber\\
&&\times \,\exp\Bigl\{ -\alpha \Bigl( {\bf k} - \alpha 
{F\over 8\gamma w^\prime} {\bf b} \Bigr)^2 \Bigr\}
\label{Z37}
\end{eqnarray}
where $\bar r_1 = a_1 + {\bf b}_1 {\bf R}$. This Green's function is positive
definite and may be viewed as a distribution function for the wave vector 
${\bf k}$. We may define the average wave vector $\langle {\bf k} 
\rangle_{\bf R}$ at space point ${\bf R}$ by 
\begin{equation}
\langle {\bf k} \rangle_{\bf R} = \int {d^dk\over (2\pi)^d} \ {\bf k} 
\  \tilde G({\bf R},{\bf k}) \,\Big/\, \int {d^dk\over (2\pi)^d} 
\ \tilde G({\bf R},{\bf k}) \ . 
\label{Z38}
\end{equation}
This formula can be rewritten in terms of the real-space Green's function 
$G({\bf r},{\bf r}^\prime)$ as 
\begin{equation}
\langle {\bf k} \rangle_{\bf R} = \bigl\{ (2i)^{-1} [\bbox{\nabla}-
\bbox{\nabla}^\prime] \, G({\bf r},{\bf r}^\prime) \bigr\} \big/
G({\bf r},{\bf r}^\prime) \big\vert_{{\bf r}^\prime={\bf r}={\bf R}} \ . 
\label{Z39}
\end{equation}
Thus, by using (\ref{Z29})-(\ref{Z32}) we obtain
\begin{equation}
\langle {\bf k} \rangle_{\bf R} = {\bf J}_{\rm s} / n_{\rm s} = {F\over 
8\gamma w^\prime} \  {{\bf b}\over -r_1} \ {(-\zeta)\, {\cal F}_{\epsilon/2}
(\zeta) \over {\cal F}_{-1+\epsilon/2}(\zeta)}
\label{Z40}
\end{equation}
so that ${\bf J}_{\rm s}= n_{\rm s} \,\langle {\bf k} \rangle_{\bf R}$. 
While ${\bf J}_{\rm s}$ is finite for $d<4$, $n_{\rm s}$ is ultraviolet 
divergent for $d>2$. As a consequence, for $d=3$ dimensions the average wave
vector $\langle {\bf k} \rangle_{\rm R}$ is strictly speaking zero. In the
superfluid state for temperatures well below $T_\lambda$ we may use the 
asymptotic formula (\ref{C44}) for the function ${\cal F}_\alpha(\zeta)$.
In this approximation the ultraviolet divergences are neglected so that we
obtain a finite result for the average wave vector 
\begin{equation}
\langle {\bf k} \rangle_{\bf R} = {F\over 8\gamma w^\prime} \  
{{\bf b}\over -r_1} \ {(-\zeta)^{3/2} \over 3^{1/2}} \ . 
\label{Z41}
\end{equation}
Since the wave vector ${\bf k}$ is related to the superfluid velocity 
${\bf v}_{\rm s}$ by ${\bf v}_{\rm s}=\hbar {\bf k}/m_4$, the Green's 
function (\ref{Z37}) implies a distribution function for the superfluid 
velocity and (\ref{Z41}) yields the average superfluid velocity $\langle
{\bf v}_{\rm s} \rangle_{\bf R}$ at space point ${\bf R}$. 

\subsection{Concluding remarks}
We have evaluated the Green's function $G_{22}=\langle\psi\psi^*\rangle$ 
only for equal times $t^\prime=t$. However, it is possible to evaluate
this Green's function also for $t^\prime \neq t$. To do this we apply the
formula (\ref{Z10}) to the factor $e^{\alpha A} e^{\beta B}$ in (\ref{Z08}).
The eventual result is again an integral over a single variable $\alpha$
with, however, a somewhat more complicated integrand than (\ref{Z28}). 
Furthermore, the nondiagonal Green's function $G_{12}=\langle \tilde\psi\psi
\rangle$ can be calculated. In this case there is only one inverse operator
$L^{-1}$ which implies only one integral over $\alpha$. This integral is 
performed trivially by a delta function analogous to (\ref{Z09}) so that 
eventually there will be no integral at all. The other nondiagonal Green's 
function $G_{21}=G_{12}^+$ is just the Hermitian conjugate. Finally, the 
upper left diagonal element $G_{11}$ is zero. Thus, we conclude that the 
complete matrix Green's function $G$ defined by (\ref{B18}) can be evaluated 
explicitly, supposed the parameters $r_1$ and $\Delta r$ are linear functions
of the space variable ${\bf r}$ according to (\ref{Z12}) and (\ref{Z13}).

\end{document}